\documentclass[longauth]{aa}  
\usepackage{graphicx}
\usepackage{txfonts}
\usepackage{enumitem}   

\usepackage[colorlinks = true, linkcolor = blue, urlcolor  = blue, citecolor = blue]{hyperref}
\usepackage{euclid}
\usepackage[switch, modulo]{lineno}

\newcommand{\orcid}[1]{} 

\def \om{{$\Omega_{\rm m} \, $}}

\begin{document} 

   \title{\Euclid preparation. XXVII. Covariance model validation for the \\2-point correlation function of galaxy clusters}

    \author{Euclid Collaboration: A.~Fumagalli$^{1,2,3,4}$\thanks{\email{alessandra.fumagalli@inaf.it}}, A.~Saro\orcid{0000-0002-9288-862X}$^{1,2,3,4}$, S.~Borgani\orcid{0000-0001-6151-6439}$^{1,3,2,4}$, T.~Castro\orcid{0000-0002-6292-3228}$^{3,2,4}$, M.~Costanzi$^{1,3,2}$, P.~Monaco$^{1,2,4,3}$, E.~Munari\orcid{0000-0002-1751-5946}$^{2}$, E.~Sefusatti\orcid{0000-0003-0473-1567}$^{2,4,3}$, N.~Aghanim$^{5}$, N.~Auricchio\orcid{0000-0003-4444-8651}$^{6}$
% ...
}

    \institute{$^{1}$ Dipartimento di Fisica - Sezione di Astronomia, Universit\'a di Trieste, Via Tiepolo 11, 34131 Trieste, Italy\\
    $^{2}$ INAF-Osservatorio Astronomico di Trieste, Via G. B. Tiepolo 11, 34143 Trieste, Italy\\
    $^{3}$ IFPU, Institute for Fundamental Physics of the Universe, via Beirut 2, 34151 Trieste, Italy\\
    $^{4}$ INFN, Sezione di Trieste, Via Valerio 2, 34127 Trieste TS, Italy\\
    $^{5}$ Universit\'e Paris-Saclay, CNRS, Institut d'astrophysique spatiale, 91405, Orsay, France\\
    $^{6}$ INAF-Osservatorio di Astrofisica e Scienza dello Spazio di Bologna, Via Piero Gobetti 93/3, 40129 Bologna, Italy\\
    $^{7}$ Dipartimento di Fisica e Astronomia "Augusto Righi" - Alma Mater Studiorum Universit\`{a} di Bologna, via Piero Gobetti 93/2, 40129 Bologna, Italy\\
    $^{8}$ INFN-Sezione di Bologna, Viale Berti Pichat 6/2, 40127 Bologna, Italy\\
    $^{9}$ Max Planck Institute for Extraterrestrial Physics, Giessenbachstr. 1, 85748 Garching, Germany\\
    $^{10}$ INAF-Osservatorio Astrofisico di Torino, Via Osservatorio 20, 10025 Pino Torinese (TO), Italy\\
    $^{11}$ Dipartimento di Fisica, Universit\`{a} di Genova, Via Dodecaneso 33, 16146, Genova, Italy\\
    $^{12}$ INFN-Sezione di Roma Tre, Via della Vasca Navale 84, 00146, Roma, Italy\\
    $^{13}$ Department of Physics "E. Pancini", University Federico II, Via Cinthia 6, 80126, Napoli, Italy\\
    $^{14}$ INAF-Osservatorio Astronomico di Capodimonte, Via Moiariello 16, 80131 Napoli, Italy\\
    $^{15}$ Instituto de Astrof\'isica e Ci\^encias do Espa\c{c}o, Universidade do Porto, CAUP, Rua das Estrelas, PT4150-762 Porto, Portugal\\
    $^{16}$ Dipartimento di Fisica, Universit\'a degli Studi di Torino, Via P. Giuria 1, 10125 Torino, Italy\\
    $^{17}$ INFN-Sezione di Torino, Via P. Giuria 1, 10125 Torino, Italy\\
    $^{18}$ INAF-IASF Milano, Via Alfonso Corti 12, 20133 Milano, Italy\\
    $^{19}$ Institut de F\'{i}sica d'Altes Energies (IFAE), The Barcelona Institute of Science and Technology, Campus UAB, 08193 Bellaterra (Barcelona), Spain\\
    $^{20}$ Port d'Informaci\'{o} Cient\'{i}fica, Campus UAB, C. Albareda s/n, 08193 Bellaterra (Barcelona), Spain\\
    $^{21}$ Institut d'Estudis Espacials de Catalunya (IEEC), Carrer Gran Capit\'a 2-4, 08034 Barcelona, Spain\\
    $^{22}$ Institute of Space Sciences (ICE, CSIC), Campus UAB, Carrer de Can Magrans, s/n, 08193 Barcelona, Spain\\
    $^{23}$ INAF-Osservatorio Astronomico di Roma, Via Frascati 33, 00078 Monteporzio Catone, Italy\\
    $^{24}$ INFN section of Naples, Via Cinthia 6, 80126, Napoli, Italy\\
    $^{25}$ Centre National d'Etudes Spatiales, Toulouse, France\\
    $^{26}$ Institut national de physique nucl\'eaire et de physique des particules, 3 rue Michel-Ange, 75794 Paris C\'edex 16, France\\
    $^{27}$ Institute for Astronomy, University of Edinburgh, Royal Observatory, Blackford Hill, Edinburgh EH9 3HJ, UK\\
    $^{28}$ Jodrell Bank Centre for Astrophysics, Department of Physics and Astronomy, University of Manchester, Oxford Road, Manchester M13 9PL, UK\\
    $^{29}$ ESAC/ESA, Camino Bajo del Castillo, s/n., Urb. Villafranca del Castillo, 28692 Villanueva de la Ca\~nada, Madrid, Spain\\
    $^{30}$ European Space Agency/ESRIN, Largo Galileo Galilei 1, 00044 Frascati, Roma, Italy\\
    $^{31}$ Univ Lyon, Univ Claude Bernard Lyon 1, CNRS/IN2P3, IP2I Lyon, UMR 5822, 69622, Villeurbanne, France\\
    $^{32}$ Institute of Physics, Laboratory of Astrophysics, Ecole Polytechnique F\'{e}d\'{e}rale de Lausanne (EPFL), Observatoire de Sauverny, 1290 Versoix, Switzerland\\
    $^{33}$ Mullard Space Science Laboratory, University College London, Holmbury St Mary, Dorking, Surrey RH5 6NT, UK\\
    $^{34}$ Departamento de F\'isica, Faculdade de Ci\^encias, Universidade de Lisboa, Edif\'icio C8, Campo Grande, PT1749-016 Lisboa, Portugal\\
    $^{35}$ Instituto de Astrof\'isica e Ci\^encias do Espa\c{c}o, Faculdade de Ci\^encias, Universidade de Lisboa, Campo Grande, 1749-016 Lisboa, Portugal\\
    $^{36}$ Department of Astronomy, University of Geneva, ch. d'Ecogia 16, CH-1290 Versoix, Switzerland\\
    $^{37}$ INFN-Padova, Via Marzolo 8, 35131 Padova, Italy\\
    $^{38}$ Universit\'e Paris-Saclay, Universit\'e Paris Cit\'e, CEA, CNRS, Astrophysique, Instrumentation et Mod\'elisation Paris-Saclay, 91191 Gif-sur-Yvette, France\\
    $^{39}$ Aix-Marseille Universit\'e, CNRS/IN2P3, CPPM, Marseille, France\\
    $^{40}$ Istituto Nazionale di Fisica Nucleare, Sezione di Bologna, Via Irnerio 46, 40126 Bologna, Italy\\
    $^{41}$ INAF-Osservatorio Astronomico di Padova, Via dell'Osservatorio 5, 35122 Padova, Italy\\
    $^{42}$ Universit\"ats-Sternwarte M\"unchen, Fakult\"at f\"ur Physik, Ludwig-Maximilians-Universit\"at M\"unchen, Scheinerstrasse 1, 81679 M\"unchen, Germany\\
    $^{43}$ Institute of Theoretical Astrophysics, University of Oslo, P.O. Box 1029 Blindern, 0315 Oslo, Norway\\
    $^{44}$ Jet Propulsion Laboratory, California Institute of Technology, 4800 Oak Grove Drive, Pasadena, CA, 91109, USA\\
    $^{45}$ Technical University of Denmark, Elektrovej 327, 2800 Kgs. Lyngby, Denmark\\
    $^{46}$ Cosmic Dawn Center (DAWN), Denmark\\
    $^{47}$ Institut d'Astrophysique de Paris, 98bis Boulevard Arago, 75014, Paris, France\\
    $^{48}$ Max-Planck-Institut f\"ur Astronomie, K\"onigstuhl 17, 69117 Heidelberg, Germany\\
    $^{49}$ Universit\'e de Gen\`eve, D\'epartement de Physique Th\'eorique and Centre for Astroparticle Physics, 24 quai Ernest-Ansermet, CH-1211 Gen\`eve 4, Switzerland\\
    $^{50}$ Department of Physics, P.O. Box 64, 00014 University of Helsinki, Finland\\
    $^{51}$ Helsinki Institute of Physics, Gustaf H{\"a}llstr{\"o}min katu 2, University of Helsinki, Helsinki, Finland\\
    $^{52}$ NOVA optical infrared instrumentation group at ASTRON, Oude Hoogeveensedijk 4, 7991PD, Dwingeloo, The Netherlands\\
    $^{53}$ Argelander-Institut f\"ur Astronomie, Universit\"at Bonn, Auf dem H\"ugel 71, 53121 Bonn, Germany\\
    $^{54}$ Department of Physics, Institute for Computational Cosmology, Durham University, South Road, DH1 3LE, UK\\
    $^{55}$ Universit\'e C\^{o}te d'Azur, Observatoire de la C\^{o}te d'Azur, CNRS, Laboratoire Lagrange, Bd de l'Observatoire, CS 34229, 06304 Nice cedex 4, France\\
    $^{56}$ European Space Agency/ESTEC, Keplerlaan 1, 2201 AZ Noordwijk, The Netherlands\\
    $^{57}$ Department of Physics and Astronomy, University of Aarhus, Ny Munkegade 120, DK-8000 Aarhus C, Denmark\\
    $^{58}$ Centre for Astrophysics, University of Waterloo, Waterloo, Ontario N2L 3G1, Canada\\
    $^{59}$ Department of Physics and Astronomy, University of Waterloo, Waterloo, Ontario N2L 3G1, Canada\\
    $^{60}$ Perimeter Institute for Theoretical Physics, Waterloo, Ontario N2L 2Y5, Canada\\
    $^{61}$ AIM, CEA, CNRS, Universit\'{e} Paris-Saclay, Universit\'{e} de Paris, 91191 Gif-sur-Yvette, France\\
    $^{62}$ Space Science Data Center, Italian Space Agency, via del Politecnico snc, 00133 Roma, Italy\\
    $^{63}$ Instituto de Astrof\'isica de Canarias, Calle V\'ia L\'actea s/n, 38204, San Crist\'obal de La Laguna, Tenerife, Spain\\
    $^{64}$ Departamento de Astrof\'{i}sica, Universidad de La Laguna, 38206, La Laguna, Tenerife, Spain\\
    $^{65}$ Dipartimento di Fisica e Astronomia "G.Galilei", Universit\'a di Padova, Via Marzolo 8, 35131 Padova, Italy\\
    $^{66}$ Departamento de F\'isica, FCFM, Universidad de Chile, Blanco Encalada 2008, Santiago, Chile\\
    $^{67}$ Centro de Investigaciones Energ\'eticas, Medioambientales y Tecnol\'ogicas (CIEMAT), Avenida Complutense 40, 28040 Madrid, Spain\\
    $^{68}$ Instituto de Astrof\'isica e Ci\^encias do Espa\c{c}o, Faculdade de Ci\^encias, Universidade de Lisboa, Tapada da Ajuda, 1349-018 Lisboa, Portugal\\
    $^{69}$ Universidad Polit\'ecnica de Cartagena, Departamento de Electr\'onica y Tecnolog\'ia de Computadoras, 30202 Cartagena, Spain\\
    $^{70}$ Institut de Recherche en Astrophysique et Plan\'etologie (IRAP), Universit\'e de Toulouse, CNRS, UPS, CNES, 14 Av. Edouard Belin, 31400 Toulouse, France\\
    $^{71}$ Infrared Processing and Analysis Center, California Institute of Technology, Pasadena, CA 91125, USA\\
    $^{72}$ INAF-Osservatorio Astronomico di Brera, Via Brera 28, 20122 Milano, Italy\\
    $^{73}$  Universit\'e Paris Cit\'e, CNRS, Astroparticule et Cosmologie, 75013 Paris, France\\
    $^{74}$ Center for Computational Astrophysics, Flatiron Institute, 162 5th Avenue, 10010, New York, NY, USA\\
    $^{75}$ School of Physics and Astronomy, Cardiff University, The Parade, Cardiff, CF24 3AA, UK\\
    $^{76}$ INAF-Istituto di Astrofisica e Planetologia Spaziali, via del Fosso del Cavaliere, 100, 00100 Roma, Italy\\
    $^{77}$ Dipartimento di Fisica "Aldo Pontremoli", Universit\'a degli Studi di Milano, Via Celoria 16, 20133 Milano, Italy\\
    $^{78}$ INFN-Sezione di Milano, Via Celoria 16, 20133 Milano, Italy\\
    $^{79}$ Dipartimento di Fisica e Astronomia "Augusto Righi" - Alma Mater Studiorum Universit\'a di Bologna, Viale Berti Pichat 6/2, 40127 Bologna, Italy\\
    $^{80}$ Junia, EPA department, 59000 Lille, France\\
    $^{81}$ SISSA, International School for Advanced Studies, Via Bonomea 265, 34136 Trieste TS, Italy\\
    $^{82}$ Dipartimento di Fisica e Scienze della Terra, Universit\'a degli Studi di Ferrara, Via Giuseppe Saragat 1, 44122 Ferrara, Italy\\
    $^{83}$ Istituto Nazionale di Fisica Nucleare, Sezione di Ferrara, Via Giuseppe Saragat 1, 44122 Ferrara, Italy\\
    $^{84}$ Institut d'Astrophysique de Paris, UMR 7095, CNRS, and Sorbonne Universit\'e, 98 bis boulevard Arago, 75014 Paris, France\\
    $^{85}$ Institut de Physique Th\'eorique, CEA, CNRS, Universit\'e Paris-Saclay 91191 Gif-sur-Yvette Cedex, France\\
    $^{86}$ NASA Ames Research Center, Moffett Field, CA 94035, USA\\
    $^{87}$ INAF, Istituto di Radioastronomia, Via Piero Gobetti 101, 40129 Bologna, Italy\\
    $^{88}$ INFN-Bologna, Via Irnerio 46, 40126 Bologna, Italy\\
    $^{89}$ Institute for Theoretical Particle Physics and Cosmology (TTK), RWTH Aachen University, 52056 Aachen, Germany\\
    $^{90}$ Institute for Astronomy, University of Hawaii, 2680 Woodlawn Drive, Honolulu, HI 96822, USA\\
    $^{91}$ Department of Physics \& Astronomy, University of California Irvine, Irvine CA 92697, USA\\
    $^{92}$ University of Lyon, UCB Lyon 1, CNRS/IN2P3, IUF, IP2I Lyon, France\\
    $^{93}$ INFN-Sezione di Genova, Via Dodecaneso 33, 16146, Genova, Italy\\
    $^{94}$ Aix-Marseille Universit\'e, CNRS, CNES, LAM, Marseille, France\\
    $^{95}$ Department of Astronomy \& Physics and Institute for Computational Astrophysics, Saint Mary's University, 923 Robie Street, Halifax, Nova Scotia, B3H 3C3, Canada\\
    $^{96}$ Department of Physics, Oxford University, Keble Road, Oxford OX1 3RH, UK\\
    $^{97}$ Instituto de F\'isica Te\'orica UAM-CSIC, Campus de Cantoblanco, 28049 Madrid, Spain\\
    $^{98}$ University Observatory, Faculty of Physics, Ludwig-Maximilians-Universit{\"a}t, Scheinerstr. 1, 81679 Munich, Germany\\
    $^{99}$ Ruhr University Bochum, Faculty of Physics and Astronomy, Astronomical Institute (AIRUB), German Centre for Cosmological Lensing (GCCL), 44780 Bochum, Germany\\
    $^{100}$ Department of Physics, Lancaster University, Lancaster, LA1 4YB, UK\\
    $^{101}$ Univ. Grenoble Alpes, CNRS, Grenoble INP, LPSC-IN2P3, 53, Avenue des Martyrs, 38000, Grenoble, France\\
    $^{102}$ Department of Physics and Astronomy, University College London, Gower Street, London WC1E 6BT, UK\\
    $^{103}$ Department of Physics and Helsinki Institute of Physics, Gustaf H\"allstr\"omin katu 2, 00014 University of Helsinki, Finland\\
    $^{104}$ Astrophysics Group, Blackett Laboratory, Imperial College London, London SW7 2AZ, UK\\
    $^{105}$ Dipartimento di Fisica, Sapienza Universit\`a di Roma, Piazzale Aldo Moro 2, 00185 Roma, Italy\\
    $^{106}$ University of Applied Sciences and Arts of Northwestern Switzerland, School of Engineering, 5210 Windisch, Switzerland\\
    $^{107}$ Institut f\"ur Theoretische Physik, University of Heidelberg, Philosophenweg 16, 69120 Heidelberg, Germany\\
    $^{108}$ Zentrum f\"ur Astronomie, Universit\"at Heidelberg, Philosophenweg 12, 69120 Heidelberg, Germany\\
    $^{109}$ Institute of Cosmology and Gravitation, University of Portsmouth, Portsmouth PO1 3FX, UK\\
    $^{110}$ Department of Mathematics and Physics E. De Giorgi, University of Salento, Via per Arnesano, CP-I93, 73100, Lecce, Italy\\
    $^{111}$ INAF-Sezione di Lecce, c/o Dipartimento Matematica e Fisica, Via per Arnesano, 73100, Lecce, Italy\\
    $^{112}$ INFN, Sezione di Lecce, Via per Arnesano, CP-193, 73100, Lecce, Italy\\
    $^{113}$ Institute of Space Science, Bucharest, 077125, Romania\\
    $^{114}$ Institute for Computational Science, University of Zurich, Winterthurerstrasse 190, 8057 Zurich, Switzerland\\
    $^{115}$ Higgs Centre for Theoretical Physics, School of Physics and Astronomy, The University of Edinburgh, Edinburgh EH9 3FD, UK\\
    $^{116}$ Universit\'e St Joseph; Faculty of Sciences, Beirut, Lebanon}
    
   \date{Received ???; accepted ???}

 
  \abstract
   {}
   {We validate a semi-analytical model for the covariance of real-space 2-point correlation function of galaxy clusters. }
   {Using 1000 PINOCCHIO light cones mimicking the expected \Euclid sample of galaxy clusters, we calibrate a simple model to accurately describe the clustering covariance. Then, we use such a model to quantify the likelihood analysis response to variations of the covariance, and investigate the impact of a cosmology-dependent matrix at the level of statistics expected for the \Euclid survey of galaxy clusters.
 }
   {We find that a Gaussian model with Poissonian shot-noise does not correctly predict the covariance of the 2-point correlation function of galaxy clusters.  By introducing few additional parameters fitted from simulations, the proposed model reproduces the numerical covariance with 10 per cent accuracy, with differences of about 5 per cent on the figure of merit of the cosmological parameters $\Omega_{\rm m}$ and $\sigma_8$. Also, we find that the cosmology-dependence of the covariance adds valuable information that is not contained in the mean value, significantly improving the constraining power of cluster clustering. Finally, we find that the cosmological figure of merit can be further improved by taking mass binning into account. Our results have significant implications for the derivation of cosmological constraints from the 2-point clustering statistics of the \Euclid survey of galaxy clusters.}
   {}


    \titlerunning{\Euclid: Cluster clustering covariance}
    \authorrunning{A. Fumagalli et al.}
   \maketitle
\section{Introduction} \label{sec:intro}
The clustering of galaxy clusters is an  increasingly powerful tool to extract cosmological information, being sensitive both to the geometry and the evolution of the large-scale structure of the Universe \citep{Borgani:1998sfa,Moscardini:1999ba,Estrada:2008em,Marulli:2018owk,Marulli:2020uyy}. Although still poorly constraining when considered alone, due to low statistic, cluster clustering is especially useful when combined with other probes, such as number counts or weak gravitational lensing, for two main reasons. First, having a different cosmology dependence, it makes it possible to break the degeneracies on parameters and improve the constraining power of these observables \citep{Schuecker:2002ti,Sereno:2014eea,Sartoris:2015aga}. Second, one of the main limitations in the cosmological exploitation of galaxy clusters lies in the fact that cluster masses have to be indirectly inferred through observable properties, such as the cluster richness, velocity dispersion, X-ray temperature or Sunyaev-Zeldovich signal. The calibration of such mass-observable scaling relations is affected by systematic biases and observational uncertainties  \citep[e.g.][]{Kravtsov:2012zs,Pratt:2019cnf}. Cluster clustering, presenting different degeneracies on parameters with respect to cluster number counts, can help to calibrate such relations, reducing the uncertainties in the mass estimation and further improving the constraints on cosmological parameters \citep{Majumdar:2003mw,Mana:2013qba,DES:2020mlx,2022arXiv220307398L}. The correlation function of galaxy clusters has also been used to identify the Baryonic Acoustic Oscillations (BAO) in a CMB-independent way \citep{Miller:2001cf, Angulo:2005pt, Huetsi:2009zq,Veropalumbo:2013cua,Moresco:2020quj}. 

The clustering of clusters presents some advantages with respect to the clustering of galaxies. Rising from the highest density peaks of the density field, galaxy clusters are a highly biased tracer of the large-scale structure, i.e., with a larger clustering signal easily detectable also at large scales. Cluster clustering can be observed on large scales, where linear theory is still suitable for describing its properties (i.e., $k \lesssim 0.05\,h\,$Mpc$^{-1}$ or $r \gtrsim 30\,h^{-1}\,$Mpc). Also, bias is primarily a function of the halo mass and can be calibrated using multi-wavelength observations. Moreover, cosmology enters in the relation between bias and mass/redshift, increasing the constraining power of cluster clustering \citep{Mo:1995cs,Tinker:2010my}. Finally, \citet{Castro:2020yes} showed that the net effect of baryons is to change the mass of clusters with negligible impact on the clustering of matched objects in dark matter and hydro simulations. Thus, we can identify the clustering of clusters as clustering of dark matter halos. For this reason, the terms ``cluster'' and ``halo'' will be used as synonyms.

The clustering of clusters is expected to become an important source of information within a few years, when upcoming and future surveys will provide  cluster samples over sizable portions of the sky. Among them, the European Space Agency (ESA) mission \Euclid,\footnote{\url{http://www.euclid-ec.org}} planned for 2023, will map $\sim\,15\,000$ deg$^2$ of the extragalactic sky, in optical and near-infrared bands, with the aim of investigating the nature of dark energy, dark matter, and gravity \citep{EUCLID:2011zbd,Euclid:2021icp}. Galaxy clusters are among the cosmological probes to be used by \Euclid, for which the mission is expected to yield a sample of  $\sim$\,$10^5$ clusters up to redshift $z \sim 2$. 
\citet{Sartoris:2015aga} showed that the constraints on cosmological parameters from cluster number counts are significantly improved when the cluster clustering information is included in the analysis of a \Euclid-like survey.

A fundamental ingredient in deriving precise and accurate constraints on cosmological parameters from these catalogs is the correct description of the uncertainties affecting the observables, which are given in the form of covariance matrices. The simplest but computationally expensive way to compute a covariance matrix is from measurements in a large set of simulations. The computational costs can be reduced by generating mocks with approximate methods instead of full N-body simulations \citep{Monaco:2016pys} or with mixed methods such as the shrinkage technique \citep{Pope:2007vz}. However, the resulting matrix will still be noisy unless a large number of mocks realizations are generated. If the covariance is considered cosmology dependent, the cost will inevitably increase as many more simulations are required to explore the high-dimensional space of cosmological parameters with such simulations.
An alternative approach is to estimate covariances from the data itself by means of bootstrap or jackknife techniques: these methods have the advantage of providing matrices evaluated at the true cosmology of the Universe, but the resampling methods tend to overestimate the true covariance, especially for two-point statistics \citep{Norberg:2008tg,Friedrich:2015nga,Lacasa:2017xbi,Mohammad:2021aqc}. A third method consists in the analytic calculation of the covariance matrix \citep[e.g.][]{Feldman:1993ky,Scoccimarro:1999kp,Meiksin:1998mu,Hu:2002we,Takada:2013wfa}, which provides noise-free, cosmology-dependent matrices without requiring expensive computational resources. The limitation of this method lies in the difficulty of describing analytically all the contributions to the covariance (e.g. non-linearities, non-Gaussianities, ...). Moreover, it is straightforward to include a treatment of systematic errors by imposing a realistic selection function to mock catalogs, while in the case of an analytical model this is more challenging and likely to require significant approximations. Therefore, such models have to be validated against simulations, in order to determine which contributions are relevant at the required level of statistics. Moreover, in some cases, this validation process may indicate that an analytical description is not sufficient to correctly describe the covariance matrix, and it is thus necessary to calibrate model parameters that cannot be determined from first principles \citep{Xu:2012hg,OConnell:2015src,Fumagalli:2022plg}.

The covariance of 2-point correlation functions is non-trivial to be modeled, due to its dependence on high-order statistics and on the survey geometry \citep{Bernstein:1993nb,Li:2018scc}. Several works have developed models for the covariance of galaxy correlation functions, both in configuration and Fourier space \citep[see, for instance,][]{Scoccimarro:1999kp,Meiksin:1998mu,Takada:2013wfa,Wadekar:2019rdu,Philcox:2019xzt,Li:2018scc}. Galaxy clustering is characterised by a Gaussian covariance, representing the main contribution at large scales, plus a non-Gaussian term arising from nonlinear gravitational instability, galaxy/halo bias and redshift-space distortions, which dominates at small scales. In addition, the coupling between short-wavelengths modes with perturbations larger than the survey size, also induced by non-Gaussianities, namely super-sample covariance, contributes to the error budget on small scales. Lastly, the shape of the observed volume can also have an impact on the covariance, requiring a convolution of the power spectrum with the window function of the survey. For cluster clustering the situation is in principle simpler, as the scales involved are larger and  mostly linear. This feature makes it possible to ignore highly non-linear effects, such as super-sample covariance, since it dominates the non-Gaussian errors in the weakly or deeply nonlinear regime \citep{Takada:2013wfa}. However, the lower densities characterising these objects produce a different weight of the various contributions \citep[e.g. shot noise, ][]{Paech:2016hod} to the covariance, compared to the case of galaxies, which could make non-Gaussian terms relevant even in the linear regime. Up to now, the analytical covariance for cluster clustering has rarely been studied \citep{Valageas:2011mz,DES:2020uce}, preferring instead numerical methods or internal estimates.

This work represents a second paper on a series, following \citet{Euclid:2021api}. We validate a semi-analytical model for the covariance of the 2-point correlation function  (2PCF hereafter) of clusters, by comparison with a numerical matrix. Since the final purpose is to apply this model covariance in the analysis of photometric data, we simply consider the real-space clustering. In fact, the redshift-space distortions of the monopole of the 2PCF are negligible with respect to the distortion produced by the photo-$z$ uncertainties \citep{Veropalumbo:2013cua,Sereno:2014eea,2022arXiv220307398L}. We test the validity of a Gaussian model, with the addition of a low-order non-Gaussian term. We are interested in understanding whether such a simple model is suitable to describe the covariance for the future survey of galaxy clusters to be extracted from the \Euclid photometric survey, estimating the impact of the missing high-order terms and of the shot-noise. Then, we focus our attention on the study of the cosmology-dependence of the covariance, to determine if this dependence can help to obtain a more precise estimate of the cosmological parameters. Lastly, we test the impact of mass binning on the cosmological constraints.
We perform the validation of the covariance for a Vanilla $\Lambda$CDM cosmological model, by studying the effect of the covariance on the cosmological constraints of the parameters $\Omega_{\rm m}$ and $\sigma_8$.

The paper is structured as follows: in Sect.\,\ref{sec:theory} we introduce the analytical formalism to describe the 2PCF and its covariance, as well as the formalism for the likelihood analysis and the estimation of the posteriors accuracy. In Sect.\,\ref{sec:simulated_data} we describe the simulated data: in Sect.\,\ref{sec:simulations} we present the simulations used for measuring the numerical matrix and for the cosmological forecasts, while in Sect\,\ref{sec:measurements} we describe the measurements of the 2PCF and the associated numerical covariance. In Sect.\,\ref{sec:results} we present the results of our analysis: in Sect.\,\ref{sec:binning} we define the best binning in spatial separation and redshift to extract the cosmological information, while in Sect.\,\ref{sec:model} we compare the analytical and numerical matrices, introducing additional parameters to improve the agreement between the two covariances. Further motivations about the need of introducing these additional parameters are discussed in Appendix\,\ref{app:shot_noise}, while the result of the fit of these parameters is detailed presented in Appendix\,\ref{app:cov_fit}. In Sect.\,\ref{sec:gauss_model} we study the impact of the non-Gaussian term, and in Sect.\,\ref{sec:cosmo_dep} we investigate the effect of the cosmology-dependent matrix; more considerations about the cosmology-dependence are shown in Appendix\,\ref{app:cosmo_mocks} and Appendix\,\ref{app:numbercounts}. In Sect.\,\ref{sec:mass_bins} we evaluate the impact of mass binning. Finally, in Sect.\,\ref{sec:conclusions} we discuss our conclusions.

\section{Theoretical background} \label{sec:theory}
In this section, we introduce the real-space 2PCF of halos and its covariance matrix model. We also describe the likelihood adopted for the parameter inference and the the method to assess the accuracy of the results.

\subsection{2-point correlation function}
We quantify the clustering of clusters with the real-space 2PCF, describing the excess number of pairs with respect to a random distribution, as a function of radial separation and redshift. Such function is defined as the Fourier transform of the halo power spectrum
\begin{equation} \label{eq:2pcf}
    \xi_{\rm h}(r,z\,|\,M) = {\overline b}^{\,2}(z\,|\,M) \int \frac{\diff k\, k^2}{2\pi^2}\, P_{\rm m}(k,z)\, j_0(kr)\,,
\end{equation}
where $P_{\rm m}(k,z)$ is the matter power spectrum, $j_0(kr)$ is the zero-order spherical Bessel function, $r$ is the comoving radial separation, and ${\overline b}(z\,|\,M)$ is the effective linear bias, i.e., the linear halo bias integrated above the mass threshold $M$
\begin{equation} \label{eq:bias}
    {\overline b} (z\,|\,M) = \frac{1}{{\overline n}(z\,|\,M)}\,\int_M^\infty \diff M' \;\frac{\diff n}{\diff M}(M',z) \,b(M',z) \,,
\end{equation}
where $\diff n/\diff M$ is the halo mass function and ${\overline n}(z\,|\,M)$ is the mean number density of objects above a mass threshold
\begin{equation} \label{eq:density}
    {\overline n}(z\,|\,M) = \int_M^\infty \diff M' \;\frac{\diff n}{\diff M}(M',z) \,.
\end{equation}
In the following, we adopt the \citet{Despali:2015yla} model to describe the halo mass function (Eq.\,7 in the paper) and the \citet{Tinker:2010my} model for the halo bias (Eq.\,6 in the paper).

For sake of simplicity, we validate our model considering halos with mass above a fixed threshold; subsequently, in Sect.\,\ref{sec:th_massbins}, we extend the discussion to the case with mass binning.

Although we work in linear theory, around the BAO scale $r_{\rm BAO} \simeq 110\,h^{-1}\,$Mpc  \citep{SDSS:2005xqv,2dFGRS:2005yhx} we account for the smoothing of the acoustic features induced by large-scale coherent flow. This produces a broadening and a shift of the BAO peak in the 2PCF \citep{Eisenstein:2006nk}, which can be modelled by the Infrared Resummation \citep{Senatore:2014via,Baldauf:2015xfa}. At the lowest order, the matter power spectrum is  corrected as
\begin{equation}
    P_{\rm m}(k,z) \simeq P_{\rm nw}(k,z)+{\rm e}^{-k^2\Sigma^2(z)}P_{\rm w}(k,z)\,,
\end{equation}
where $P_{\rm w}$ and $P_{\rm nw}$ are, respectively, the wiggle and non-wiggle parts of the linear power spectrum, and
\begin{equation}
        \Sigma^2(z) = \int_0^{k_s}\frac{\diff q}{6\pi^2}\, P_{\rm nw}(q,z)\,\left[1-j_0\left(q\,r_{\rm BAO}\right)+2j_2\left(q\,r_{\rm BAO}\right)\right]\,.
\end{equation}

The final expression for the real-space 2PCF of halos to be compared with observations is obtained by averaging Eq.\,\eqref{eq:2pcf}  over the $a$-th redshift bin and $i$-th separation bin,
\begin{equation}
    \xi_{\rm h}^{\,a i} =  \int \frac{\diff k \, k^2}{2 \pi^2}\, \left \langle\, \overline{b}\,\sqrt{P_{\rm m}(k)}\, \right \rangle_a^2  W_i(k) \,,  
    \label{eq:2pcf_binned}   
\end{equation}
where $\langle\,\rangle_a$ indicates the average over the redshift bin:
\begin{equation}
     \left \langle\, \overline{b}\,\sqrt{P_{\rm m}(k)}\, \right \rangle_a =   \frac{\int_{\Delta z_a} \diff z\, \frac{\diff V}{\diff z }\, {\overline n}(z)\, {\overline b}(z) \sqrt{P_{\rm m}(k,z)} }{\int_{\Delta z_a} \diff z\, \frac{\diff V}{\diff z }\, {\overline n}(z)}\,,
\end{equation}
where $\diff V/\diff z = \Omega_{\rm sky} \, \diff V/\diff \Omega\,\diff z$ is the comoving volume per unit redshift and $\Omega_{\rm sky}$ is the survey area in steradians.\footnote{This expression is valid for a conical geometry survey; in more generic cases, the integral over the lightcone volume must take into account the geometry of the survey.} $W_i(k)$ represents the spherical shell window function, given by
\begin{equation}
     W_i(k) = \int \frac{\diff^3r}{V_i} j_0(kr) = \frac{r_{i,+}^3 W_{\rm th}(k r_{i,+}) - r_{i,-}^3 W_{\rm th}(k r_{i,-})}{r_{i,+}^3 - r_{i,-}^3}\,,
\end{equation}
where $W_{\rm th}(kr)$ is the top-hat window function, $V_i$ is the volume of the $i$-th spherical shell, and $r_{i,-},\,r_{i,+}$ are the extremes of the separation bin.

\subsection{Covariance model}

The 2PCF covariance can be obtained as the Fourier transform of the power spectrum covariance. The latter is defined as
\begin{equation} \label{eq:cov_def}
    C_P(\mathbf{k}, \mathbf{k'}) = \left \langle \left [\hat{P}(\mathbf{k}) - \langle \hat{P}(\mathbf{k}) \rangle \right] \, \left[ \hat{P}(\mathbf{k'}) - \langle \hat{P}(\mathbf{k'}) \rangle \right] \right \rangle\,,
\end{equation}
where 
\begin{equation}
    \hat{P}(\mathbf{k}) = V\,|\delta_\mathbf{k}|^2 - \frac{1}{{\overline n}}
\end{equation}
is the estimator for the halo power spectrum, such that $\langle \hat{P}(\mathbf{k}) \rangle = P_{\rm h}(\mathbf{k})$. Here, $V$ is the observed volume and $1/{\overline n}$ is the (Poissonian) shot-noise correction to the halo power spectrum $P_{\rm h}$. 

Substituting the power spectrum estimator in Eq.\,\eqref{eq:cov_def} we obtain the expression of the power spectrum covariance  \citep{Meiksin:1998mu,Scoccimarro:1999kp}
\begin{equation} \label{eq:pk_cov}
    \begin{split}
        C_P(\mathbf{k},& \mathbf{k'}) = \frac{(2 \pi)^3}{V} \left [ P_{\rm h}(k) + \frac{1}{{\overline n}} \right ]^2 \left[ \delta^{\rm D}(\mathbf{k}-\mathbf{k'}) +  \delta^{\rm D}(\mathbf{k}+\mathbf{k'})\right] \\
        & + \frac{1}{V\,{\overline n}^2} \bigg [ P_{\rm h}(|\mathbf{k}-\mathbf{k'}|) + P_{\rm h}(|\mathbf{k}+\mathbf{k'}|) + 2P_{\rm h}(k) + 2P_{\rm h}(k') \bigg ] \\
        & + \frac{1}{V\,{\overline n}}  \bigg [ B_{\rm h}(\mathbf{k},-\mathbf{k},0) + B_{\rm h}(0,\mathbf{k'},-\mathbf{k})+ B_{\rm h}(\mathbf{k}+\mathbf{k'}, -\mathbf{k}, -\mathbf{k'})\\
        & + B_{\rm h}(\mathbf{k}-\mathbf{k'},-\mathbf{k},\mathbf{k'}) + B_{\rm h}(\mathbf{k},\mathbf{k'}-\mathbf{k},-\mathbf{k'})+ B_{\rm h}(\mathbf{k}, -\mathbf{k'}, \mathbf{k'}) \bigg ] \\
        & + \frac{1}{V} \, {T_{\rm h}(\mathbf{k},\mathbf{-k},\mathbf{k'},\mathbf{-k'})} + \frac{1}{V \, {\overline n}^3},
    \end{split}
\end{equation}
where $B_{\rm h}$ and $T_{\rm h}$ are, respectively, the bispectrum and the trispectrum of halos, i.e., the three and four-point correlation functions in Fourier space. The first line represents the Gaussian covariance, while the other lines represent the non-Gaussian component. As motivated in Sect.\,\ref{sec:intro}, we do not consider the super-sample covariance.

By Fourier transforming Eq.\,\eqref{eq:pk_cov} and integrating over separation and redshift bins \citep{Cohn:2005ex}, we obtain a model for the 2PCF covariance in the light cone
\begin{equation} \label{eq:covariance}
    \begin{split}
        C_{a i j} =  & \, \frac{2}{V_a} \int \frac{\diff k\,k^2}{2 \pi^2} \left[ \left \langle {\overline b\,}^2P_{\rm m}(k) \right \rangle_a + \left \langle \frac{1}{{\overline n}}\right \rangle_a \right]^2 W_i(k) \, W_j(k)\\
        & + \frac{2}{V_a V_i} \int \frac{\diff k\,k^2}{2 \pi^2} \left \langle {\overline b\,}^2 P_{\rm m}(k)\right \rangle_a \left \langle\frac{1}{{\overline n}}\right \rangle^2_a \, W_j(k) \ \delta_{ij}\,,
    \end{split}
\end{equation}
where $i,j$ states for the two separation bins, while the $a$ index is for the average over the redshift bin, and $V_a$ is the volume of the redshift slice. The model in Eq.\,\eqref{eq:covariance} is clearly a simplification of the full covariance matrix based on the following approximations:
\begin{itemize}
    \item by considering large redshift slices  ($\Delta z \gtrsim 0.2$), we assume the cross-correlation between redshift bins to be negligible, as verified from the numerical matrix;
    \item we neglect the contribution from higher-order correlation functions, only including the lowest order shot-noise contributions of the non-Gaussian covariance, in addition to the Gaussian part;
    \item we do not include the terms that contribute only at zero separation ($\propto \delta^{\rm D}(r_i), \delta^{\rm D}(r_j)$), since we consider larger scales;
    \item we do not account for the survey footprint, but consider a simplistic window-function described by a fixed size opening angle.
\end{itemize}

\subsection{Mass binning}
\label{sec:th_massbins}
We now extend the formalism to take into account the mass binning instead of a simple mass threshold, in order to quantify the amount of information contained in the mass-dependence of the halo bias.

We rewrite Eq.\,\eqref{eq:2pcf} as
\begin{equation} \label{eq:2pcf_mass}
    \xi_{\rm h}(r,z\,|\,M,M') = {\overline b}(z\,|\,M)\,{\overline b}(z\,|\,M') \int \frac{\diff k\, k^2}{2\pi^2}\, P_{\rm m}(k,z)\, j_0(kr)\,,
\end{equation}
and all the equations derived in Sect.\,\ref{sec:theory} are modified according to this change. We obtain the binned 2PCF by integrating Eq.\,\eqref{eq:2pcf_mass} over the $i$-th separation bin, $a$-th redshift bin, and between $m$-th and $n$-th mass bins. Note that now the integrals over mass (Eqs.\,\ref{eq:bias} and \ref{eq:density}) are performed between the edges of each mass bin. The final binned 2PCF takes into account both the auto-correlation inside a single mass interval, and the cross-correlation between halos belonging to two different mass bins
\begin{equation}\label{eq:2pcf_binned_mass} 
    \xi_{\rm h}^{\,aimn}  =  \int \frac{\diff k \, k^2}{2 \pi^2}\, \left \langle\, \overline{b}_m\,\sqrt{P_{\rm m}(k)}\, \right \rangle_a \left \langle\, \overline{b}_n\,\sqrt{P_{\rm m}(k)}\, \right \rangle_a  W_i(k) \,.    
\end{equation}

Consequently, the covariance matrix is adapted to account for four kind of terms: auto-correlation between auto-2PCFs ($\propto \xi_{mm}^2$), auto-correlation between cross-2PCFs ($\propto \xi_{mn}^2$), cross-correlation between auto-2PCFs ($\propto \xi_{mm}\,\xi_{nn}$), and cross-correlation between cross-2PCFs ($\propto \xi_{mn}\,\xi_{pq}$), and reads as
\begin{equation}
    C_{aijmnpq} = \frac{C'_{aijmpnq} +C'_{aijmqnp}}{2}\,,
\end{equation}
where
\begin{equation}
    \label{eq:covariance_mass}
    \begin{split}
     C'_{aijmnpq}& =  \frac{2}{V_a} \int \frac{\diff k\,k^2}{2 \pi^2}  \left[ \left \langle \,{\overline b}_m\,{\overline b}_n\,P_{\rm m}(k) \right \rangle_a + \left \langle \frac{\delta_{mn}}{{\overline n_m}} \right \rangle_a \right]\\
     & \hspace{1.8cm} \times  \left[ \left \langle \,{\overline b}_p\,{\overline b}_q\,P_{\rm m}(k) \right \rangle_a + \left \langle \frac{\delta_{pq}}{{\overline n_p}} \,  \right \rangle_a \right] \,W_i\,W_j\\
    & +  \frac{2}{V_a} \int \frac{\diff k\,k^2}{2 \pi^2}  \left \langle \,{\overline b}_m\,{\overline b}_p\, P_{\rm m}(k)\right \rangle_a \left \langle\frac{\delta_{mn}}{{\overline n_m}}\,\right \rangle_a\,\left \langle\frac{\delta_{pq}}{{\overline n_p}}\,\right \rangle_a\frac{W_j}{V_i} \delta_{ij}\,.
\end{split}
\end{equation} 

\subsection{Likelihood function}
We study the effect of the covariance on cosmological constraints by performing Bayesian inference on mock cluster surveys extracted from simulations. We explore the posterior distribution with a Monte Carlo Markov Chains (MCMC) approach, by using a \textit{python} wrapper for the nested sampling \texttt{PyMultiNest}  \citep{Buchner:2014nha}. 

We adopt a Gaussian likelihood
\begin{equation} \label{gauss_cov_l}
    \mathcal{L}(\mathbf{d} \,\vert\,\mathbf{m}(\boldsymbol{\theta}),\,C) = \frac{ \exp {\left \{-\frac{1}{2} [\mathbf{d} - \mathbf{m}(\boldsymbol{\theta})]^T C^{-1}  [\mathbf{d} - \mathbf{m}(\boldsymbol{\theta})] \right \}}}{\sqrt{(2\pi)^N \vert C \vert}} \,,
\end{equation}
where $\mathbf{d}$ is the data vector, $\mathbf{m}(\boldsymbol{\theta})$ is the prediction as a function of a set of cosmological parameters $\boldsymbol{\theta}$, and $C$ is the covariance matrix, which may also depend on cosmological parameters.

To remove the effect of cosmic variance affecting the single realization of the Universe, we perform the MCMC analysis by maximizing the log-likelihood function averaged over all the $N_{\rm S}$ simulated catalogs used in this work
\begin{equation}     \label{eq:mean_logl}
    \ln \mathcal{L}^{{\rm tot}} = \frac{1}{N_{\rm S}} \sum_{a=1}^{N_{\rm S}} \ln \mathcal{L}^{(a)} \,.
\end{equation}
Thanks to this step, we are able to individuate possible systematic effects in the analysis, that could introduce small, but sizeable shifts in the cosmological posteriors with respect to the input  parameter values.

We quantify the accuracy of our covariance (error) estimates in terms of the effect on the figure of merit \citep[FoM,][]{Albrecht:2006um} for two parameters $\theta_1$ and $\theta_2$, defined as
\begin{equation} \label{eq:fom}
\mathrm{FoM}(\theta_1, \theta_2) = \vert \,\mathrm{C}(\theta_1,\theta_2) \,\vert ^{-1/2} \,,
\end{equation}
where $C(\theta_1,\theta_2)$ is the parameter covariance computed from the sampled posteriors. The FoM is proportional to the inverse of the area enclosed by the 68 per cent confidence level ellipse; therefore, in general, a higher FoM indicates more accurate evaluation of parameters. For the covariance comparison, however, a larger FoM could indicate an underestimation of the posteriors amplitude, resulting from a wrong estimation of the uncertainties on the statistical quantities entering the likelihood. We should therefore point out that we are not interested in the absolute value of the FoM, but rather in the difference between the various cases.

\section{Simulated data}
\label{sec:simulated_data}
We describe in this section the simulations used and the procedure to measure the 2PCF and its numerical covariance.

\subsection{Simulations} \label{sec:simulations}
Following \citet{Euclid:2021api}, we validate our model by comparing it with a reference covariance, computed numerically from simulations. The use of a large set of simulations is a fundamental requirement for the accurate estimation of covariance matrices. The dimension of such set depends on the size of the data vector (i.e. the total number of bins) and the desired accuracy. Typically, the required number of catalogs is around $10^3$ or even more \citep{Taylor:2012kz,Dodelson:2013uaa}. For this purpose, catalogues generated with N-body simulations are hardly obtainable, due to the high computational cost. Instead, large sets of mock data can be produced in a simpler and faster way by using approximate methods based on perturbative theories. Although less accurate than full N-body simulations in reproducing the observables, these methods are able to accurately estimate covariances requiring fewer resources and far less computational time \citep{Sahni:1995rm,Monaco:2016pys,Lippich:2018wrx,Blot:2018oxk,Colavincenzo:2018cgf}.

In this work we use a set of mock catalogs produced with the PINOCCHIO \citep[PINpointing Orbit-Crossing Collapsed HIerarchical Objects, ][]{Monaco:2001jg,Munari:2016aut} algorithm. PINOCCHIO generates dark matter halo catalogs using Lagrangian Perturbation Theory \citep[LPT,][]{Moutarde:1991evx,Buchert:1992ya,Bouchet:1994xp} up to third order and the ellipsoidal collapse \citep{Bond:1993we,Eisenstein:1994ni}.
The code generates an initial density field on a regular grid, with periodic boundary conditions, and computes the collapse time of each particle. Then, by means of LPT, it displaces particles to form halos, which are lastly moved to their final positions by applying again LPT. In this way, the code is able to simulate large cubic boxes, that are used to build the past-light cones. The latter are generated by replicating the periodic boxes through an “on-the-fly” process, selecting only the halos causally connected with an observer at the present time. 

Our data set consists of 1000 past-light cones\footnote{The light cones can be obtained on request. The list of the available mocks can be found at \url{http://adlibitum.oats.inaf.it/monaco/mocks.html}; the light cones analyzed are the ones labeled “NewClusterMocks”.}  each covering an area of 10\,313 deg$^2$ and redshift range $z=0$\,--\,2.5. \footnote{Note that our light cones are covering slightly smaller areas than the expected \Euclid catalogs ($\sim 10\,000$ vs. $\sim 15\,000$\,deg$^2$); also, the survey will cover two separate patches of the sky. For the purpose of this work, we expect that these differences impact the results in a negligible way.} The light cones contain halos with virial masses above $3.61 \times 10^{13}\,M_\odot $, sampled with more than 50 particles. The cosmology used in the simulations is the flat $\Lambda$CDM one with parameters fixed according to \citet{Planck:2013pxb} (Table 5, ``Planck+WP+highL+BAO'' case): $\Omega_{\rm m } = 0.30711$ for the total matter density parameter, $\Omega_{\rm b} = 0.048254$ for the corresponding contribution from baryons, $h = 0.6777$ for the Hubble parameter expressed in units of 100 km\,s$^{-1}$\,Mpc$^{-1}$,  $n_{\rm s} = 0.96$ for the primordial spectral index,  $A_{\rm s} = 2.21 \times 10^{-9}$ for the power spectrum normalization, and $\sigma_8 = 0.8288$ for the RMS density fluctuation at $z=0$ within a top-hat sphere of 8\,$h^{-1}$\,Mpc radius. 

To avoid complications linked to the modeling of the halo mass function, we use a version of these catalogs with masses rescaled according to the \citet{Despali:2015yla} mass function. The rescaling process has been performed by matching the average mass distribution with the predicted halo mass function, maintaining all the fluctuations due to shot-noise and sample variance in each catalog. Also, the \citet{Tinker:2010my} prediction for the halo bias has been verified to be in agreement within 10\,\% with our final catalogs, down to 5\,\% at low masses. More details about this rescaling can be found in \citet{Euclid:2021api}. The final light cones contain $\sim 10^5$ halos, each with virial mass $M_{\rm vir} \ge 10^{14}\,M_\odot$, and redshift range $z = 0$\,--\,2. In this first step, we do not include any selection function or mass-observable relation; such quantities will be added in the next stages of the analysis.

\subsection{Measurements} \label{sec:measurements}

We consider radial separations in range $r = 20$\,--\,130\,$h^{-1}$\,Mpc. Such interval includes linear scales, where the bias is almost constant \citep{Manera:2009ak}, plus the BAO peak. We consider all the halos above the mass threshold $M_{\rm vir} = 10^{14}\,M_\odot$, but it is straightforward to generalize the measurement formalism for the mass binning case.

To measure the 2PCF from simulations we use the \citet{Landy:1993yu} estimator
\begin{equation}
    \hat{\xi}_{\rm h}^{\,a i} = \frac{{\rm DD}_{ai} - 2{\rm DR}_{ai} + {\rm RR}_{ai}}{{\rm RR}_{ai}}\,,
\end{equation}
where ${\rm DD}_{ai},\,{\rm DR}_{ai},\,{\rm RR}_{ai}$ are the number of pairs in the data-data, data-random and random-random catalogs, respectively, within the $a$-th redshift bin and $i$-th separation bin, normalized for the number of objects in the data and random catalogs,$n_{\rm R}$ and $n_{\rm D}$ \citep[see, e.g.,][]{Kerscher:1999hc}.
The random catalog has been built by randomly extracting objects from each mock and stacking them together, to obtain a catalog with $\mathbf{n_{\rm R} = 10\,n_{\rm D}}$ objects randomly distributed inside the light cone volume. The measurement of the correlation function is performed with the \texttt{CosmoBolognaLib} package \citep{Marulli:2015jil}. 

\begin{figure}
    \centering
    \includegraphics[scale=0.53]{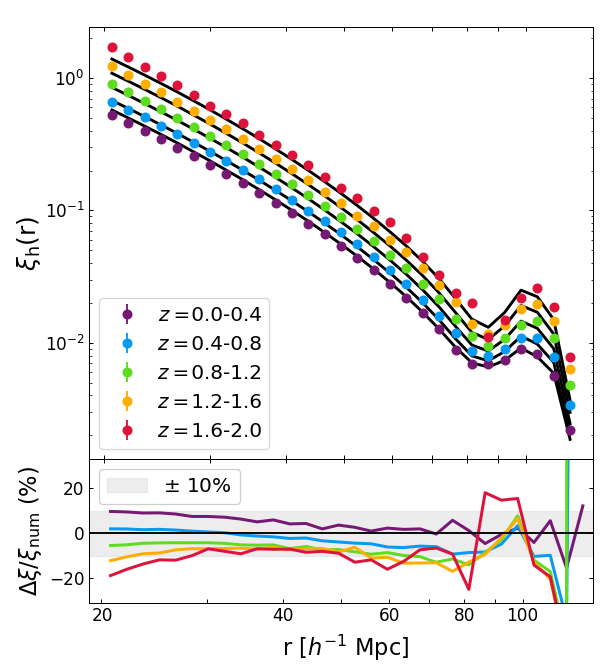}
    \caption{2PCF of halos. \textit{Top panel:} measured (colored dots) and predicted (black lines) 2PCF as a function of the radial separation, for different redshift bins. \textit{Bottom panel:}  percent residuals of the model with respect to the numerical function.}
    \label{fig:2pcf}
\end{figure}

In Fig.\,\ref{fig:2pcf}, we show the measured 2PCF in different redshift bins, as a function of the radial separation, averaged over the 1000 mocks and compared with the analytical prediction of Eq.\,\eqref{eq:2pcf_binned}. We associate to the average measured quantities an uncertainty given by the standard error on the mean, which is extremely small and thus not visible in the figure. The predicted 2PCF shows an agreement within 10\,\% with the numerical one at almost all the separations and redshifts. The differences between the various redshift bins are ascribed to the non-perfect description of the halo bias, that is underestimated at high redshift and overestimated at low redshift. 
Such difference turns out to shift the cosmological posteriors with respect to the fiducial cosmology, indicating that an accurate description of the halo bias is fundamental to obtain unbiased constraints from the cluster clustering. Since the calibration of the halo bias is beyond the purpose of this paper, we simply compensate this inaccuracy by correcting the prediction for the 2PCF in the likelihood analysis with 

\begin{equation}
    \label{eq:2pcf_correction}
    \xi'_{\rm h}(\boldsymbol{\theta})  = \xi_{\rm h}(\boldsymbol{\theta}) \frac{\langle\,\hat{\xi}_{\rm h}\,\rangle}{\xi_{\rm h}(\boldsymbol{\theta}_{\rm input})}\,,
\end{equation} 
where $\langle\,\hat{\xi}_{\rm h}\,\rangle$ is the measured 2PCF averaged over the 1000 simulations, and $\boldsymbol{\theta}_{\rm input}$ are the input parameters of the simulations. In this way, by construction, we provide an unbiased description of the 2PCF which contains the correct cosmology dependence. 

In Fig.\,\ref{fig:2pcf}, we can also notice a smaller additional difference both at small separations and around the BAO scale, due to some non-linear effects. This confirms the correct choice of the radial range, which cannot be further extended to avoid introducing errors due to the limitations of a linear model.

To compute the numerical covariance matrix, we use the estimator
\begin{equation} \label{eq:num_covariance}
    \hat{C}_{a b i j} = \frac{1}{N_{\rm S}-1} \sum_{s=1}^{N_{\rm S}} \left(\hat{\xi}_{i a}^{\,(s)} - \langle\,\hat{\xi}\,\rangle_{i a}\right) \, \left(\hat{\xi}_{j b}^{\,(s)} - \langle\,\hat{\xi}\, \rangle_{j b}\right) \,,
\end{equation}
where $N_{\rm S}$ is the number of catalogs, $\hat{\xi}_{i a}^{\,(s)}$ is the 2PCF in the $a$-th redshift bin and $i$-th radial bin measured from the $s$-th mock, and $\langle\,\hat{\xi}\,\rangle_{i a}$ is the corresponding average value. 
The uncertainty on the numerical covariance is given by \citep[see, e.g.,][]{Taylor:2012kz}
\begin{equation}
    \sigma^2(\hat{C}_{a b i j}) = \frac{1}{N_{\rm S}-1} \bigg (\hat{C}_{abij}^2 + \hat{C}_{aaii}\,\hat{C}_{bbjj} \bigg )\,.
\end{equation}

\begin{figure}
    \centering
    \includegraphics[scale=0.57]{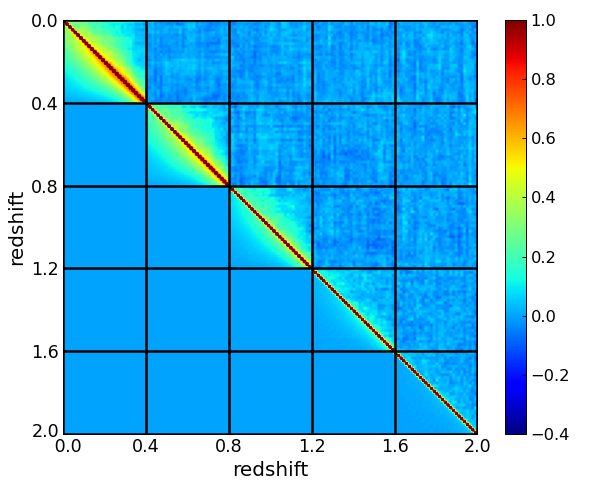}
    \caption{Numerical (upper triangle) and analytical (lower triangle) correlation matrices. Color bar is shown on the right.}
    \label{fig:matrix}
\end{figure}

In the upper triangle of Fig.\,\ref{fig:matrix}, we show the numerical correlation matrix, namely the covariance of Eq.\,\eqref{eq:num_covariance} normalised by the diagonal elements
\begin{equation}
    R_{a b i j} = \frac{C_{a b i j}}{\sqrt{\; C_{a b i i}\; C_{a b j j}\;}} \,.
\end{equation}
The result confirms the validity of the assumption of negligible cross-correlation between redshift bins, since the off-block diagonal terms of the matrix are only populated by noise consistent with zero signal. On the contrary, inside each redshift bin there is a significant non-diagonal correlation, especially at low redshift.

Due to the inaccuracy arising from the finite number of simulations, the inverse of such matrix requires to be corrected as \citep{Anderson:2003,Hartlap:2006kj}
\begin{equation}
    \hat{C}^{-1}_{\rm unbiased} = \frac{N_{\rm S}-N_{\rm D}-2}{N_{\rm S} -1} \,\hat{C}^{-1} \,,
\end{equation}
where $N_{\rm D}$ is the dimension of the data vector. As detailed in Sect.\,\ref{sec:binning}, our baseline analysis considers 5 redshift bins and 30 radial bins, i.e. $N_{\rm D} = 150$, which with $N_{\rm S}=1000$ gives a correction to the inverse matrix by a factor of $\sim 0.85$.

While this correction removes the bias in the numerical covariance, sampling noise propagates to the parameter covariance inducing an increase of errorbars by a factor \citep{Taylor:2012kz,Dodelson:2013uaa}
\begin{equation}
    f=1+\frac{(N_{\rm S}-N_{\rm D}-2)}{(N_{\rm S}-N_{\rm D}-1)(N_{\rm S}-N_{\rm D}-4)} (N_{\rm D}-N_{\rm P})\,,
\end{equation}
where $N_{\rm P}$ is the number of (cosmological + nuisance) parameters, that we take here to be $N_{\rm P}=2$. We obtain $f=1.17$, implying a 17\,\% increase of parameter errorbars due to sampling noise. To reduce this impact below 10\,\% one should nearly double the number of mocks; $N_S=2000$ would give $f=1.08$. This constraint on the number of mocks does not apply if the numerical covariance is fit with a model \citep{Fumagalli:2022plg}, as described in Sect.\,\ref{sec:model}.  
This correction results from a frequentist style approach concerned with results after repeated trials; for corrections suitable for a Bayesian analysis the proper approach is described by the works of \citet{Sellentin:2015waz,Percival:2021cuq}. However, we simply attenuate the propagation of the sampling noise on the parameter constraints by manually setting to zero the cross-correlation between redshift bins in the numerical covariance, being dominated entirely by noise. By doing so, the number of noise-affected bins in the matrix is reduced to $N_{\rm D} \sim N_{\rm r} = 30$, where $N_{\rm r}$ is the number of radial bins, providing a correction factor for the inverse covariance of $\sim 0.97$ and a negligible increase of the parameter errorbars ($f \sim 1.03$), allowing us to take the numerical results as reference for the model comparison.

\section{Results} \label{sec:results}
In this section we present the results of the covariance comparison and the effect that different covariance configurations have on the cosmological posteriors. In  Sect.\,\ref{sec:binning}, we analyse the redshift and radial binning schemes to determine the configuration that better extracts the information in the likelihood analysis; we then use such configuration for the covariance model validation. In Sect.\,\ref{sec:model}, we validate the analytical model and in Sect.\,\ref{sec:gauss_model} we study the impact of the non-Gaussian term on the covariance. Lastly, in Sect.\,\ref{sec:cosmo_dep} we study the cosmology dependence of the covariance and in Sect.\,\ref{sec:mass_bins} we evaluate the impact of the mass binning.

For the likelihood analysis we consider the cosmological parameters on which the cluster clustering is more sensitive, i.e., $\Omega_{\rm m}$ and $\sigma_8$, or equivalently $A_s$. We assume flat uninformative priors $\Omega_{\rm m} \in [0.2,\,0.4]$ and $\logten A_s \in [-9.0,\,-8.0]$, and then we derive the value of $\sigma_8$ through the relation $P_{\rm m}(k) = A_s\,k^{n_s}\,T^2(k)$, where $T(k)$ is the transfer function, and the definition of variance $\sigma^2(R)$. We are interested in evaluating the variations in the FoM in the $\Omega_{\rm m}$\,--\,$\sigma_8$ plane and the possible biases in the posteriors with respect to the input cosmology.

\subsection{Radial and redshift binning} \label{sec:binning}

Before starting the model validation, we are interested in defining the best binning scheme, to properly extract the cosmological information. To this purpose, we perform the likelihood analysis with different combinations of radial and redshift bin widths. For this test, we consider only the covariance matrix extracted from numerical simulations, i.e. the reference covariance.

We divide the separation range in different number of bins: $N_{\rm r} = 20, 25, 30, 35$ log-spaced, plus $N_{\rm r} = 25$ linearly spaced, to test the effect of a different spacing. For the redshift binning, we test three bin widths, $\Delta z = 0.2, 0.4, 0.5$, which properly divide the whole redshift range. We do not consider thicker bins, to avoid the inclusion of non-negligible border effects in the pair count procedure. 

\begin{figure}
    \centering
    \includegraphics[width=0.49\textwidth]{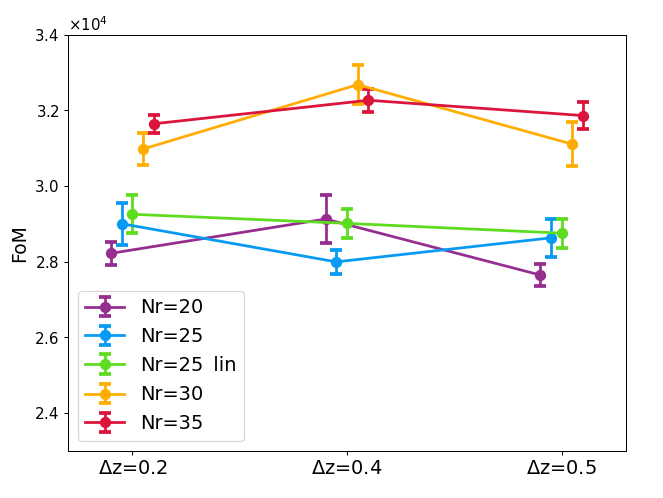}
    \caption{Figure of merit in the \om--\,$\sigma_8$ plane for different numbers of radial bins, as a function of the redshift bin width. A small horizontal displacement has been applied to make the comparison clearer.}
    \label{fig:fom_binning}
\end{figure}

In Fig.\,\ref{fig:fom_binning}, we show the FoM for the different number of radial bins, as a function of the redshift bin width. To take into account the uncertainty in the inference process, we consider the average and the standard error computed over 5 realizations for each case. We do not observe a significant difference between the various $\Delta z$, since all the cases are statistically in agreement. About the radial binning, the FoM increases as the number of bins increases, suggesting a more efficient extraction of the information, and stabilises around $N_{\rm r} = 30$, meaning that no more information can be extracted by further increasing the number of radial bins.

In the following analyses, we adopt the values $\Delta z = 0.4$ and $N_{\rm r} = 30$ log-spaced as our baseline redshift and radial bins choice.

\subsection{Covariance comparison} \label{sec:model}

\begin{figure*}
    \centering
    \includegraphics[width=1.01\textwidth]{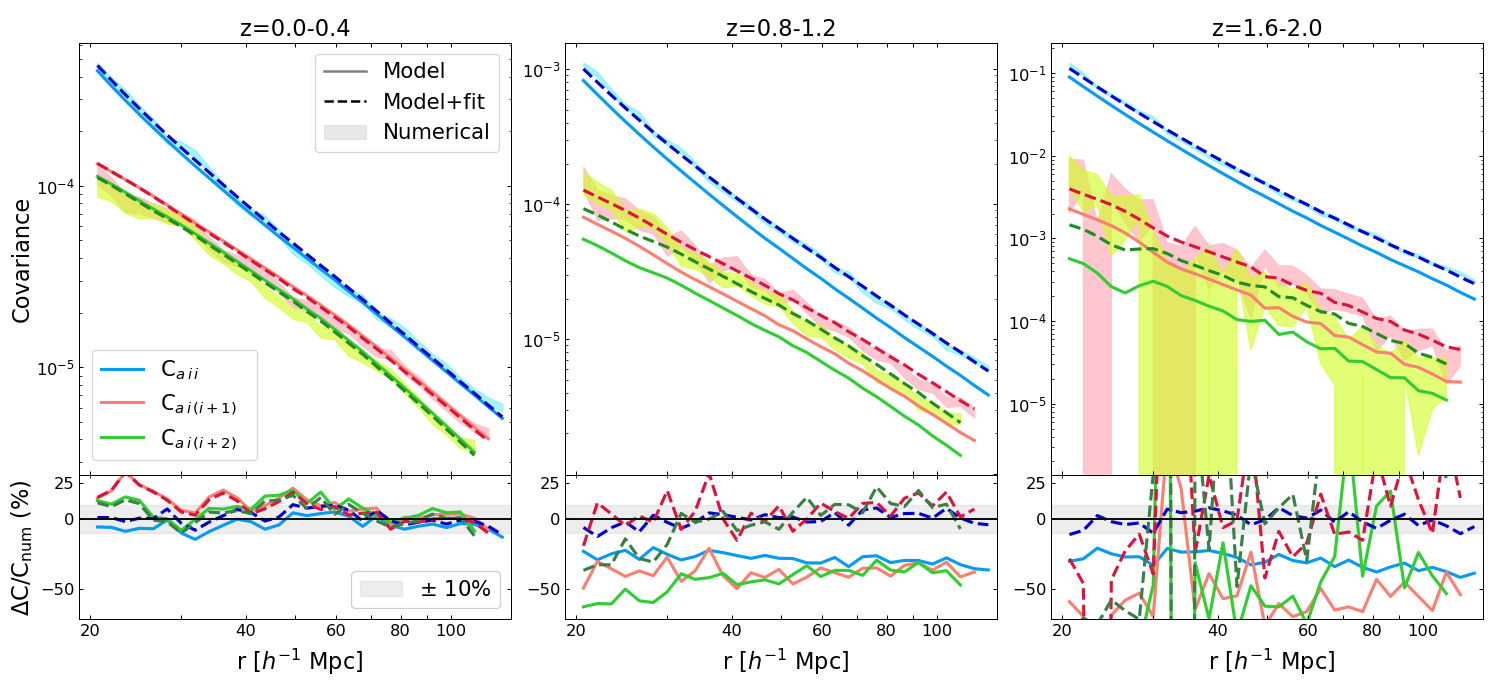}
    \caption{Numerical (shaded areas),  analytical (solid lines), and analytical with fitted parameters (dashed lines) covariance matrices as a function of the radial separation, in three redshift bins (from the left to the right panels: $z=0.0$\,--\,0.4, $z=0.8$\,--\,1.2, $z=1.6$\,--\,2.0). Different colors represent different components of the matrix: diagonal elements in blue, first off-diagonal elements in red and second off-diagonal elements in green. In the subpanels, percent residuals of the model covariance with respect the numerical matrix.}
    \label{fig:covariance}
\end{figure*}

\begin{figure}
    \centering\includegraphics[width=0.49\textwidth]{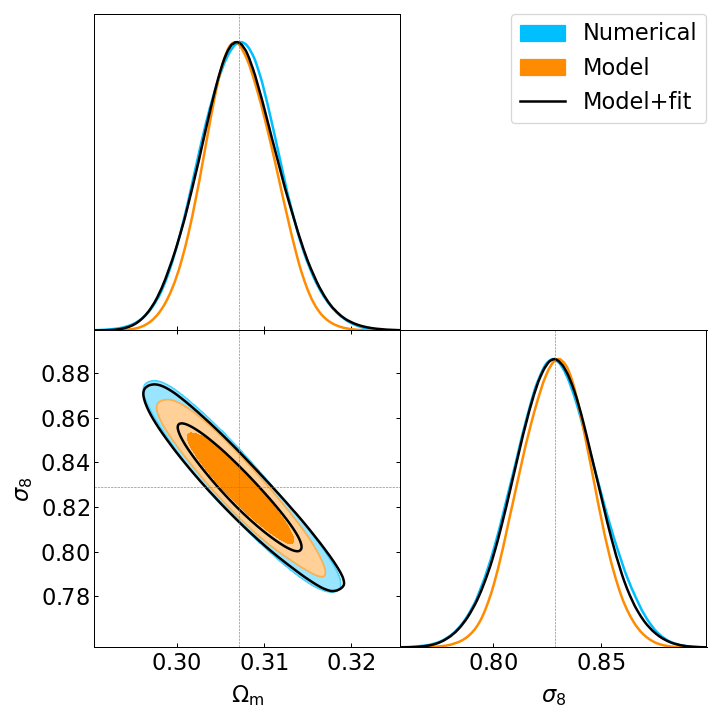}
    \caption{Contour plots at 68 and 95 per cent of confidence level for the numerical (blue), analytical (orange) and analytical with fitted parameters (black) matrices. Dotted gray lines represent the fiducial cosmology.}
    \label{fig:post_model}
\end{figure}

In this section, we present the validation of the analytical model of Eq.\,\eqref{eq:covariance}, through the comparison with the numerical matrix. The two correlation matrices are represented, respectively, in the lower and upper triangle of Fig.\,\ref{fig:matrix}. For a better comparison, in Fig.\,\ref{fig:covariance} we show the diagonal and two off-diagonal terms of the matrices as a function of the radial separation, in three redshift bins. We can see that the model (solid lines) correctly reproduces the reference values (shaded areas) only at low redshift, while at intermediate and high redshift it underestimates the numerical matrix by about 30\,\% on the diagonal and by about 50\,\% on the off-diagonal terms. We ascribe this difference to three factors:
\begin{itemize}
    \item non-Poissonian shot noise: the Poissonian prediction does not properly describe the shot-noise affecting the halo power spectrum (see Appendix \ref{app:shot_noise} for further discussion);
    \item inaccurate halo bias: the inaccuracy of the halo bias prediction propagates in the covariance model; \footnote{Note that the correction of Eq.\,\eqref{eq:2pcf_correction} does not apply to the covariance prediction. However, this does not affect the following results, as we treat the bias in the 2PCF and the one in the covariance model as two different quantities.}
    \item lack of higher-order terms: the contribution of tri- and four-point functions is not negligible. This effect especially regards the terms weighted by $1/{\overline n}$, that would give a significant contribution at high redshift, where the shot noise increases.
\end{itemize}

We correct the inaccuracy of the predicted covariance by including some parameters in the model. More specifically, we modify Eq.\,\eqref{eq:covariance} by adding three free parameters $\{\alpha,\beta,\gamma\}$
\begin{equation} \label{eq:covariance_fit}
    \begin{split}
    C_{a i j} & =  \, \frac{2}{V_a} \int \frac{\diff k\,k^2}{2 \pi^2} \left[ \left \langle ({\beta \, \overline b\,})^2P_{\rm m}(k) \right \rangle_a + \left \langle \frac{1 + \alpha}{{\overline n}}\right \rangle_a \right]^2 W_i(k) \, W_j(k)\\
    & + \frac{2}{V_a V_i} \int \frac{\diff k\,k^2}{2 \pi^2} \left \langle ({\beta \, \overline b\,})^2 P_{\rm m}(k)\right \rangle_a \left \langle\frac{1+\gamma}{{\overline n}}\right \rangle^2_a \, W_j(k) \ \delta_{ij}\,,
\end{split}
\end{equation}
where $\beta$ corrects for the halo bias inaccuracy, and $\alpha$ and $\gamma$ correct for the non-Poissonian nature of the shot-noise in the main and secondary term, respectively; the different weighting of the shot-noise correction should also account for the effect of higher-order terms. We fit such parameters from simulations in each redshift bin, assuming a constant value with scale and redshift in each slice. We adopted the method described in  \citet{Fumagalli:2022plg} to fit the free parameters $\alpha$, $\beta$, and $\gamma$. In short, we constrain the covariance through the maximization of a Gaussian likelihood evaluated at the fiducial cosmology, with free covariance parameters. The best-fit covariance thus obtained is the one that best follows a $\chi^2$ distribution with respect to the observed data (we refer to the original paper and to Appendix\,\ref{app:cov_fit} for more details on the method). In Table\,\ref{table:params_fit}, we show the best-fit values of the parameters in each redshift slice\footnote{We show the value of best-fit parameters for general considerations. However, the value of these parameters is not universal, but depends on the properties of the survey and must be fitted for each specific case.}: in most of the cases the best-fit is not in agreement with the reference values. The correction of the halo bias is in line with the expectation (i.e., $\beta < 1$ at low redshift to correct an overestimated bias, and $\beta > 1$ at high redshift to correct an underestimated bias); anyway, at redshift $z \gtrsim 1$, the values of $\beta$ overestimate the 2PCF correction of Eq.\,\eqref{eq:2pcf_correction} by a factor from 5 to 30\,\% depending on redshift. The shot-noise corrections also show conflicting results: the Gaussian term of the covariance seems to prefer a super-Poissonian shot-noise ($\alpha > 0$), while the non-Gaussian term is characterized by a sub-Poissonian shot-noise ($\gamma < 0$). Such contrasts suggest that the parameters actually absorb the effect of the wrong or missing terms of the covariance, instead of simply describing the halo bias correction or the deviation from the Poissonian prediction of the shot-noise.

The dashed lines in Fig.\,\ref{fig:covariance} show the predictions of the model modified by the introduction of the additional parameters. Now, the analytical covariance correctly describes the numerical results at all redshifts, with an accuracy of about 10 per cent.

In Fig.\,\ref{fig:post_model}, we show the posterior distributions resulting from the likelihood analysis with three different covariance configurations: numerical, model of Eq.\,\eqref{eq:covariance} and model of Eq.\,\eqref{eq:covariance_fit} with the best-fit parameters shown in Table\,\ref{table:params_fit}. As expected, the underestimated level of covariance provided by the original model translates in tighter posteriors with respect to the numerical case. On the contrary, the model corrected for the additional parameters recovers with good accuracy the result of the numerical matrix. The FoM obtained from these posteriors and the percent difference with respect to the numerical case are shown in Table\,\ref{table:fom}: the addition of parameters decreases the deviation in the FoM from  $\sim$ 40\,\% to only $\sim$ 5\,\%.

\begin{table}[t]
\centering           
\caption{Best-fit values for the covariance model parameters introduced in Eq\,\eqref{eq:covariance_fit}.}
\begin{tabular}{l c c c}     
\hline  
Redshift & $\alpha$ & $\beta$ & $\gamma$ \\
\hline
0.0 -- 0.4  & 0.111 $\pm$ 0.008 & 0.979 $\pm$ 0.008 & $-$0.027 $\pm$ 0.047\\
0.4 -- 0.8  & 0.109 $\pm$ 0.008 & 1.055 $\pm$ 0.009 & $-$0.083 $\pm$ 0.037 \\
0.8 -- 1.2  & 0.134 $\pm$ 0.008 & 1.181 $\pm$ 0.013 & $-$0.129 $\pm$ 0.027 \\
1.2 -- 1.6  & 0.157 $\pm$ 0.008 & 1.270 $\pm$ 0.022 & $-$0.199 $\pm$ 0.024 \\
1.6 -- 2.0  & 0.188 $\pm$ 0.008 & 1.460 $\pm$ 0.045 & $-$0.263 $\pm$ 0.026 \\
\hline
Reference   & 0  &  1  &  0 \\
\hline
\label{table:params_fit}    
\end{tabular}
\end{table}

\begin{table}[t]
\centering           
\caption{Figure of merit for the different covariance cases. In the third column, percent difference with respect to the numerical case.}
\begin{tabular}{l c c}       
\hline  
Case & FoM & $\Delta$FoM\;/\;FoM$_{\rm num}$ \\
\hline
Numerical    & 32681 $\pm$ 514 & -- \\
Model        & 45510 $\pm$ 413 & + 39\,\% \\
Model + fit  & 34307 $\pm$ 623 & + 5\ \% \\
Model + fit, Gauss & 38855 $\pm$ 437 & + 19\,\% \\
Cosmo-dependent & 86155 $\pm$  670 &  + 151\,\% \\
\hline
\label{table:fom} 
\end{tabular}
\end{table}

\subsection{Non-Gaussian term} \label{sec:gauss_model}
\begin{figure}
    \centering
    \includegraphics[scale=0.5]{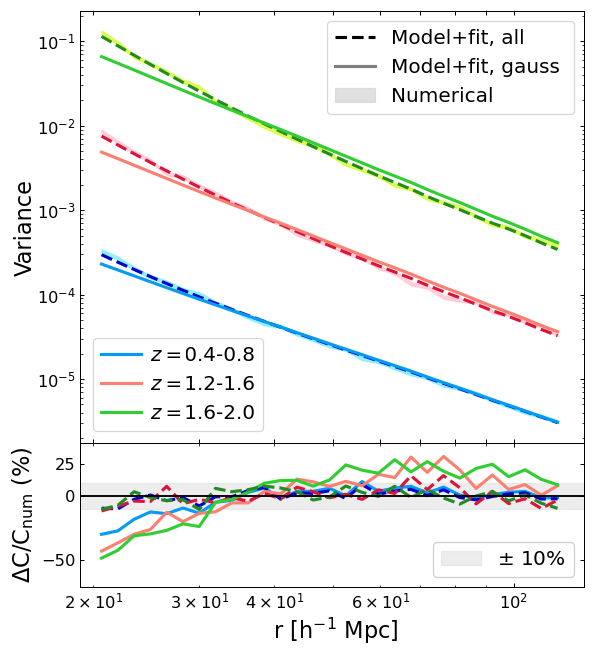}
    \caption{Variance as a function of the radial separation for three redshift bins, for the numerical matrix (shaded area), Gaussian analytical matrix (solid lines) and full analytical matrix (dashed lines, corresponding to the dashed lines of Fig.\,\ref{fig:covariance}). }
    \label{fig:cov_gauss}
\end{figure}

We test here the effect of the low-order non-Gaussian term (i.e., second line in Eq.\,\ref{eq:covariance}), to evaluate its impact with respect to the Gaussian covariance. In Fig.\,\ref{fig:cov_gauss} we compare the numerical matrix with the analytical model, with parameters fitted both from the full model (dashed lines), and from the Gaussian model, i.e., setting to zero the non-Gaussian term and fitting $\alpha$ and $\beta$ (solid lines). Since such a term only contributes on the diagonal elements, we compare the variance in three different redshift bins. The figure clearly shows that the Gaussian model is unable to properly describe the numerical covariance, for two reasons: first, the non-Gaussian term gives a significant contribution at small scales, especially at high redshift, and neglecting this term leads to an underestimation of the diagonal terms by a factor up to 50 per cent. Second, the Gaussian model does not have enough degrees of freedom to provide a good fit and is not able to absorb the effect of the missing terms, thus producing a wrong fit also at larger scales. 

The differences in the Gaussian fit have an impact on the cosmological posteriors, with deviations in the FoM of about 20 per cent with respect to the numerical covariance case (see Table\,\ref{table:fom}).

Since the importance of this term is mainly driven by the factor $\overline{n}^{\,-2}$ that grows with decreasing number of objects, we expect that the impact of this term increases when considering higher redshifts, as well as higher mass-limits. The same trend would apply to the bispectrum terms, due to the factor $\overline{n}^{\,-1}$, while the trispectum contribution should be less relevant, given the absence of such a factor.

\subsection{Cosmology dependence} \label{sec:cosmo_dep}

\begin{figure*}[t!]
    \centering
    \includegraphics[width=1.01\textwidth]{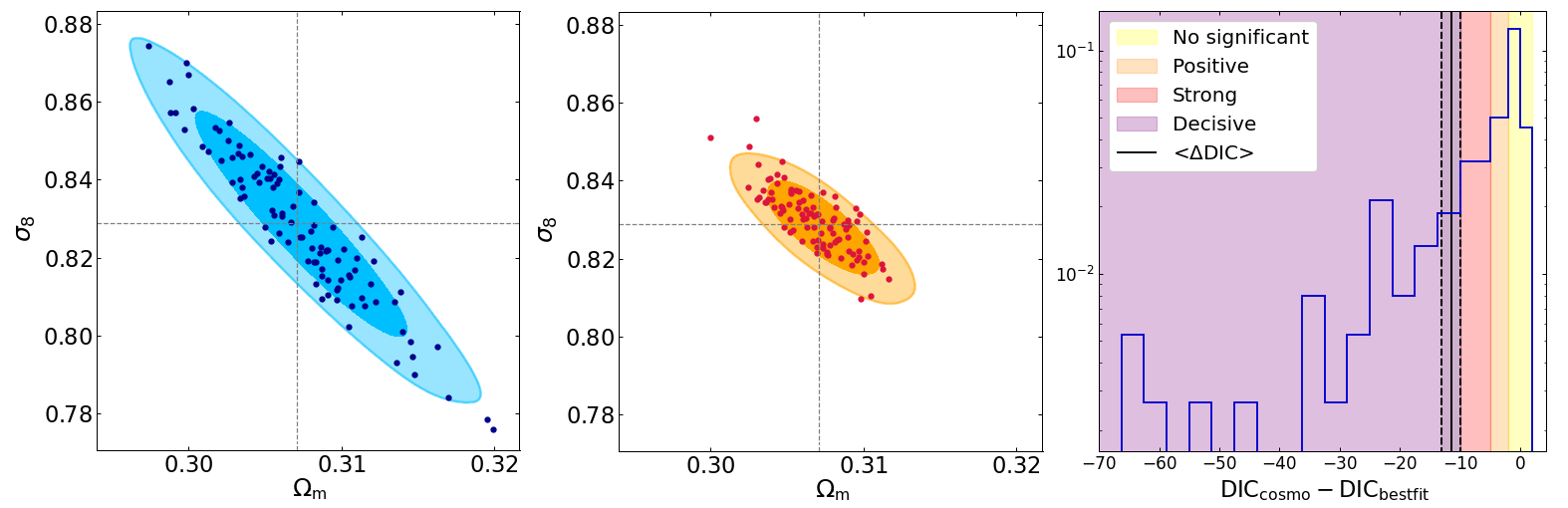}
    \caption{\textit{Left and middle panels:} Contour plots at 68 and 95 per cent of confidence level for input-cosmology covariance (blue), and the cosmology-dependent covariance (orange), obtained from the mean likelihood (Eq.\,\ref{eq:mean_logl}). Dots are the best-fit values from 100 single light cones. Gray lines represent the input parameters. \textit{Right panel:} $\Delta$DIC distribution of the 100 light cones. Associated mean and error on the mean are highlighted as solid and dashed black lines. Colored regions represent the Jefferys’ scale used to interpret the results.} 
    \label{fig:post_hist_dic}
\end{figure*}

\begin{figure}
    \centering
    \includegraphics[width=0.49\textwidth]{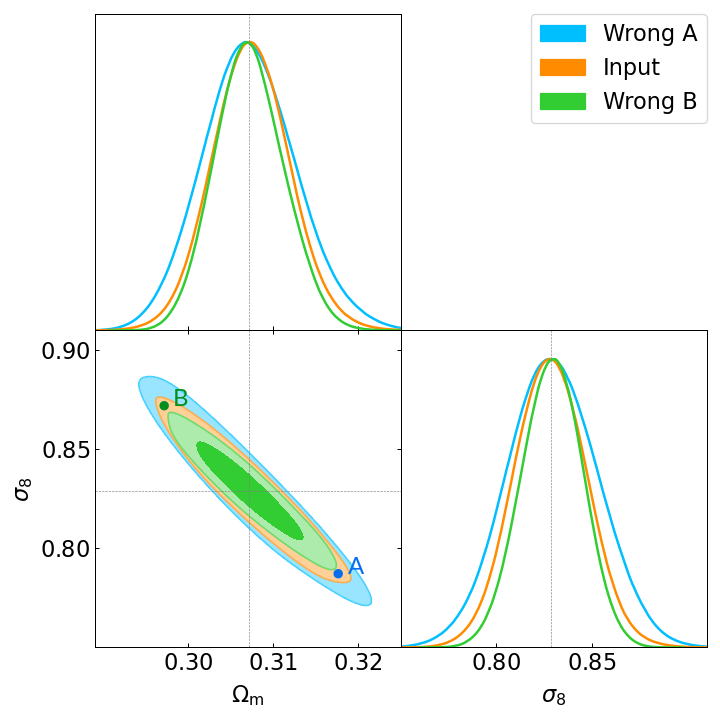}
    \caption{Contour plots at 68 and 95 per cent of confidence level for covariance matrix at the input cosmology (orange), compared with two wrong-cosmology cases: $\Omega_{\rm m}=0.320,\, \sigma_8=0.775$ in blue, and $\Omega_{\rm m}=0.295,\,\sigma_8=0.871$ in green (cases A and B respectively). The colored dots indicate the position of the wrong parameters. Dotted gray lines represent the fiducial cosmology.}
    \label{fig:post_wrong_cosmo}
\end{figure}

\begin{figure}
    \centering
    \includegraphics[width=0.49\textwidth]{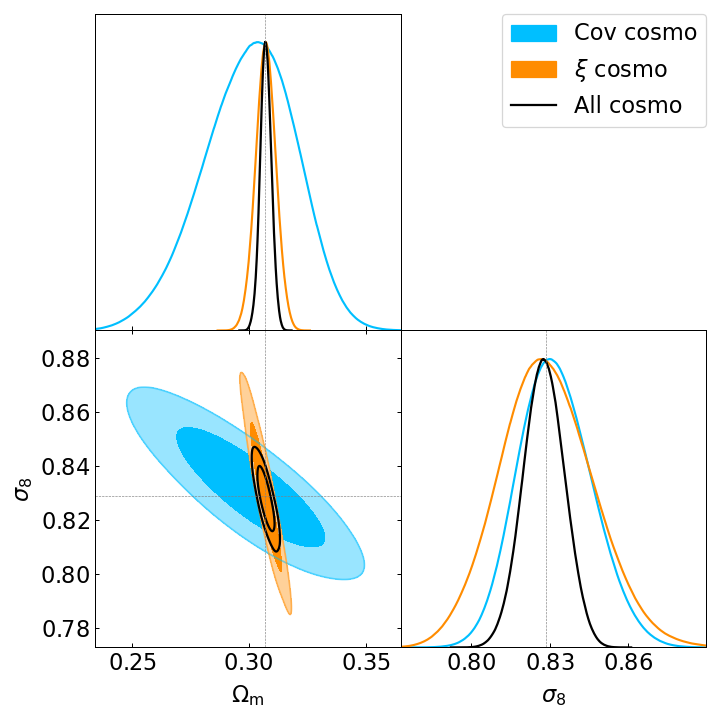}
    \caption{Contour plots at 68 and 95 per cent of confidence level for three cases: cosmology-dependent matrix and fixed mean value (blue), fixed covariance and cosmology-dependent mean value (orange), and cosmology-dependent mean value and covariance (black). Dotted gray lines represent the fiducial cosmology.}
    \label{fig:post_cosmo}
\end{figure}

The impact of the cosmology dependence of the covariance in the likelihood analysis is a topic that has been largely discussed in literature. Several works \citep[e.g.][]{Krause:2016jvl,Eifler:2008gx,Morrison:2013tqa,Blot:2020cbi,Euclid:2021api} have demonstrated that evaluating the covariance matrix at a wrong cosmology would lead to a wrong estimation of the cosmological posteriors. To avoid this, the correct way to perform the parameter inference from a Gaussian likelihood is to use a cosmology-dependent covariance, i.e., to recompute the matrix at each step of the MCMC process. The situation gets more complicated if the Gaussian likelihood is just an approximation of the true distribution of the data, as for the 2PCF. As pointed out by \citet{Carron:2012pw}, in this case the use of a cosmology-dependent covariance may lead to a wrong estimation of the posteriors amplitude. To avoid this, one can use the iterative approach, which consists in running the MCMC with a fixed covariance computed at some fiducial cosmology, then using the best-fit parameters to construct a new covariance matrix and re-running the MCMC process. This can be iterated until convergence of the cosmological posteriors. It should be noted that, in the case of approximate likelihood, even this second method may not correctly estimate the posteriors amplitude.

Given this premise, we perform the following test to establish which is the most correct method to extract the cosmological information from the 2PCF, with the likelihood and the covariance model proposed in this work. We analyse 100 light cones in two different ways: i) we apply the iterative method starting with a fiducial cosmology of $\Omega_{\rm m} = 0.30$ and $\sigma_8 = 0.77$, and verifying that a single step is sufficient to achieve convergence; ii) we use a cosmology-dependent covariance. The left and middle panels of Fig.\,\ref{fig:post_hist_dic} represent the result of the two analysis: dots are the best fit values for each light cone, compared to the mean contours obtained through Eq.\,\eqref{eq:mean_logl}. We can see the two cases exhibit a different best-fit distribution: analyzing the light cones with the cosmology-dependent covariance yields values that are more concentrated around the input cosmology, compared to the fixed covariance case, which instead presents a more scattered distribution. In both cases the individual values are in agreement with the mean distribution, making it difficult to determine which of the two analyses is more correct. Thus, for a better comparison, we compute the  Deviance Information Criterion \citep[DIC,][]{Spiegelhalter:2002yvw}, treating the problem as a model selection problem. The DIC is defined as
\begin{equation}
    {\rm DIC}(m_i) = \langle \chi^2 \rangle + p_D\,,
\end{equation}
with
\begin{equation}
    p_D = \langle \chi^2 \rangle - \chi^2(\boldsymbol{\theta}_{\rm input})\,.
\end{equation}
Here $\chi^2 = -2 \ln \mathcal{L}(\mathbf{d}| m_i(\boldsymbol{\theta}),C)$ estimates the goodness of the fit and $p_D$ is the Bayesian complexity, measuring the effective complexity of the model. The average is performed over the posteriors volume. Given two models $m_1(\boldsymbol{\theta})$ and $m_2(\boldsymbol{\theta})$, the difference $\Delta {\rm DIC} = {\rm DIC}(m_2) - {\rm DIC}(m_1)$ is interpreted using the Jeffreys’ scale presented in \citet{Grandis:2016fwl}: $\Delta {\rm DIC} = 0$ means that none of the two models is preferred, $ -2 < \Delta {\rm DIC} < 0$ that there is ``no significant'' preference for $m_2$, $ -5 < \Delta {\rm DIC} < -2$ a ``positive'' preference for $m_2$, $ -10 < \Delta {\rm DIC} < -5$ a ``strong'' preference for $m_2$, and $\Delta {\rm DIC} < -10$ indicates a ``decisive'' preference for $m_2$. By defining $\Delta {\rm DIC} = {\rm DIC}_{\rm cosmo} - {\rm DIC}_{\rm bestfit}$ for each of the 100 simulations, we obtain the distribution shown in the right panel of Fig.\,\ref{fig:post_hist_dic}, characterised by a mean value $\langle \Delta {\rm DIC} \rangle_{\rm sims} = -11.5 \pm 1.6$.  The analysis of the $\Delta$DIC indicates that the model with cosmology-dependent covariance is statistically preferred over the iterative method. Further considerations are presented in Appendix\,\ref{app:cosmo_mocks}.

After verifying that the use of the cosmology-dependence covariance is, from a statistical point of view, the most correct way to analyse the data, we study the impact on the (average) cosmological posteriors of a wrong-cosmology covariance  and a cosmology-dependent covariance, with respect to the input covariance case.
In Fig.\,\ref{fig:post_wrong_cosmo}, we show the posteriors obtained by fixing the covariance matrix at three different cosmologies. More specifically, we compare the input-parameter case ($\Omega_{\rm m}=0.307,\, \sigma_8=0.829$) with two choices of parameter combinations, i.e. $\Omega_{\rm m}=0.320,\, \sigma_8=0.775$ and $\Omega_{\rm m}=0.295,\,\sigma_8=0.871$, located approximately at the extremes of the 2$\sigma$ contours of the input-cosmology posteriors, along the degeneracy direction (indicated by dots in the figure, with respect to the orange contours). Note that such deviations from the fiducial cosmology are comparable with the 2$\sigma$ values from \citet{Planck:2018vyg}, which represents the state of the art in the cosmological constraints.  We observe that using the covariance matrix computed at a wrong cosmology has a significative effect on the cosmological posteriors, with variations in the FoM of the order of $\sim 30\,-\,40\,\%$. We note that the recovered posterior distributions differ even if the two adopted cosmologies lie along the \om--\,$\sigma_8$ degeneracies. This result suggests that the cosmological dependence of the covariance matrix is different from that of the 2PCF. 

To test this hypothesis we compare the derived posterior distribution on the cosmological parameters for the following three analyses:
\begin{enumerate}[label=\textit{\roman*)}]
    \item  We compute the covariance at the input cosmology and evaluate the expected 2PCF as a function of cosmological parameters. This case corresponds to the standard likelihood analysis with fixed covariance, where all the cosmological information is encapsulated in the expected value of $\xi(r,z)$.
    \item We evaluate the expected 2PCF at the fixed input cosmology, but let the covariance matrix vary as a function of cosmological parameters. In this way we evaluate the cosmology dependence of the covariance alone;
    \item We compare the measured and expected clustering signal where both the mean value and its covariance matrix are varying as a function of cosmological parameters. This case corresponds to the full forward-modelling approach.
\end{enumerate}

When adopting a cosmology-dependent covariance, we assume the fitted parameters $\alpha$, $\beta$, and $\gamma$ to be cosmology-independent. This limitation is due to the fact that we only have simulations for one cosmology on which perform the fit. The impact of such dependency will be verified in detail in future analyses. However, we expect that neglecting the cosmology dependence of these parameters would introduce a negligible error with respect to the one that we would introduce by not including these parameters at all. 

Figure\,\ref{fig:post_cosmo} clearly highlights a tilted degeneracy direction between $\Omega_{\rm m}$ and $\sigma_8$ posteriors of cases $i)$ and $ii$), indicating that covariance and 2PCF have a different cosmological dependencies (blue versus orange contours). As a result, by varying the cosmological parameters in both the quantities returns tighter constraints, with a FoM improved by about $150$ per cent with respect to the numerical case, which reflects the standard case $i$) likelihood analysis (see Table\,\ref{table:fom}).

This different dependence on cosmology can be explained by noting that, unlike the mean value, the covariance of the 2PCF depends on the shot-noise, which is proportional to the inverse of the integrated mass function. Letting the cosmology vary also in the covariance thus makes it possible to extract all the information contained in the clustering of the clusters, and not only in the 2PCF itself. Further considerations about the cosmology-dependence of the covariance are presented in Appendix\,\ref{app:numbercounts}.

\subsection{Mass binning}\label{sec:mass_bins}

\begin{figure}[t]
    \centering
    \includegraphics[width=0.49\textwidth]{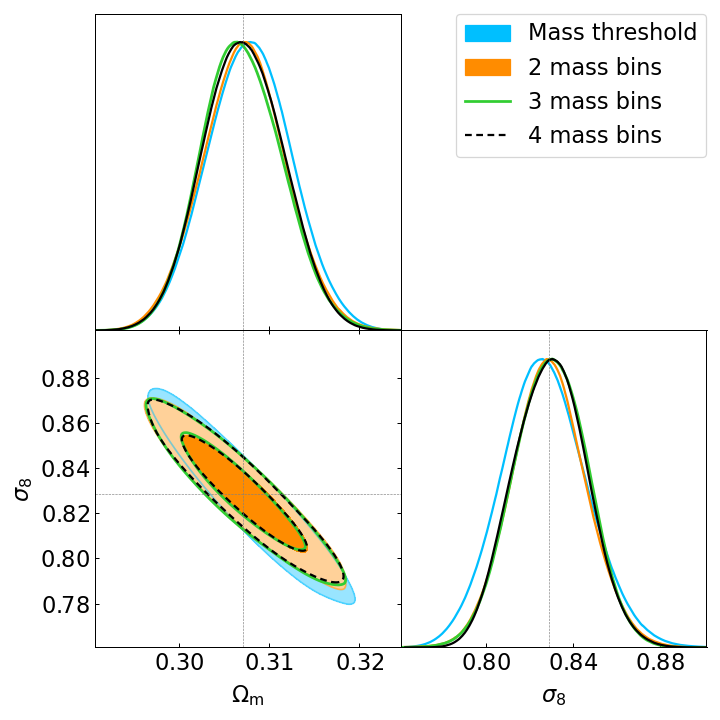}
    \caption{Contour plots at 68 and 95 per cent of confidence level for three cases: no mass binning (blue), two mass bins (orange), and three mass bins (black). In all the cases the covariance is given by the numerical matrix. Dotted gray lines represent the fiducial cosmology. }
    \label{fig:post_mass}
\end{figure}

\begin{figure*}[h]
    \centering
    \includegraphics[width=0.8\textwidth]{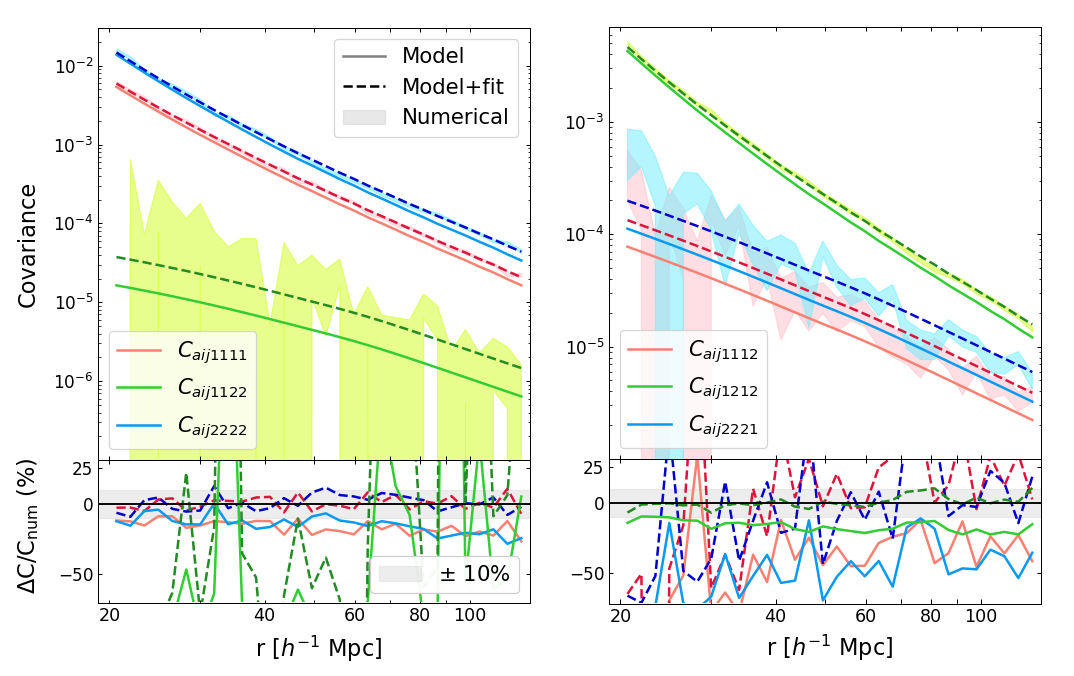}
    \caption{Numerical (shaded areas),  analytical (solid lines), and analytical with fitted parameters (dashed lines) covariance matrices as a function of the radial separation, in the redshift bin $z = 1.0\,$--\,1.5. Different colors represent the diagonal elements of different auto- and cross-correlation components of the matrix.}
    \label{fig:covariance_mass}
\end{figure*}

We perform this analysis considering redshift bins of width $\Delta z = 0.5$, to allow for more populated mass bins. We consider the case of two mass bins with cuts at $\logten M/M_\odot = \{14.00, 14.15, 16.00\}$, three mass bins with cuts at $\logten M/M_\odot = \{14.00, 14.05, 14.15, 16.00\}$ and four mass bins with cuts at $\logten M/M_\odot = \{14.00, 14.05, 14.10, 14.20, 16.00\}$, chosen in order to have at least 4000 objects in each mass and redshift bin.

We show in Fig.\,\ref{fig:post_mass} the posterior distribution from the three cases with mass binning, compared to the mass threshold case, while in Table\,\ref{table:fom_mass} we report the corresponding FoMs.
We observe an improvement in the FoM when considering the mass binning with respect of the mass-threshold case, indicating that the information included in the halo bias can be exploited in order to obtain tighter constraints on cosmological parameters. However, increasing the number of mass bins does not improve significantly the contours: this can be attributed to the closeness of the bins, characterised by a similar bias relation. On the other hand, selecting more distant bins implies having less populated intervals and therefore noisier quantities.

\begin{table}[t]
\centering           
\caption{Figure of merit for the different mass binning cases. In the third column, percent difference with respect to the ``Mass threshold'' case in the upper part, and ``2 mass bins'' numerical case in the lower one.}
\begin{tabular}{l c c}       
\hline  
Case & FoM & $\Delta$FoM\;/\;FoM$_{\rm num}$ \\
\hline
Mass threshold  & 29759 $\pm$ 554 & -- \\
2 mass bins     & 36555 $\pm$ 349 & + 23\,\% \\
3 mass bins     & 35243 $\pm$ 308 & + 18\,\% \\
4 mass bins     & 37160 $\pm$ 497 & + 25\,\% \\
\hline
Model           &  48500 $\pm$ 738 & + 33\,\% \\
Model + fit     &  37980 $\pm$ 543 & + 4\,\% \\
Cosmo-dependent & 121921 $\pm$ 615 & + 230\,\% \\
\hline
\label{table:fom_mass} 
\end{tabular}
\end{table}

Once we have established the advantage of considering mass binning, we validate the corresponding covariance model presented in Eq.\,\eqref{eq:covariance_mass}. For greater clarity, we consider the simplest two mass bins case; the cases with more mass bins are analogous. In Fig.\,\ref{fig:covariance_mass} we show the diagonal components of the analytical matrix (solid lines), compared to the corresponding numerical terms (shaded areas). As in the mass threshold case, the model underestimates the expected covariance, with a difference in the FoM of about the 30\,\% (see Table\,\ref{table:fom_mass}). 

Again, we correct this discrepancy by adding some covariance parameters, fitted for each mass and redshift bin, according to Eq.\,\eqref{eq:covariance_fit}. When adding the parameters fitted from simulations, the discrepancy between numerical and analytical matrix drops to less than 5\,\% on the FoM.

Finally, we test the effect of the cosmology-dependent covariance, following the analyses described in Sect.\,\ref{sec:cosmo_dep}. In this case, the improvement in the cosmological posteriors is even higher than the mass threshold case, reaching a difference in the FoM of $\sim $ \,230\,\%. This is due the mass-dependence of the shot-noise, that makes the covariance more constraining than the single mass threshold case.

\section{Discussion and conclusions} \label{sec:conclusions}
In this work we validate a covariance model for the real-space 2-point correlation function of galaxy clusters in a survey that is comparable to that expected from the \Euclid survey in terms of mass selection, sky coverage and depth. As this represents a first step in a more complex analysis, we do not account here for the effect of selection functions and mass-observable relations. 

We consider a Gaussian model plus the low-order non-Gaussian contribution, neglecting high-order terms. This choice is motivated since we expect the non-Gaussian terms to be minor corrections to the main Gaussian covariance. As such, great efforts to analytically calculate these complicated terms are not computationally justified. With this premise, we are interested in evaluating the impact of the approximations we made to compute this simple model, i.e. the absence of three- and four-point correlation functions, at the level of accuracy required for the future \Euclid cluster catalogs.

We perform the validation of the covariance model by comparison with a numerical matrix, estimated by means of 1000 \Euclid-like light cones generated with the PINOCCHIO algorithm.

We measure the 2PCF from the light cones with the \citet{Landy:1993yu} estimator and compare the result with the theoretical prediction of Eq.\,\eqref{eq:2pcf_binned}, in the redshift range $z = 0\,$--\,2 and radial range $r = 20\,$--\,130\,$h^{-1}$\,Mpc. We consider, in first place, halos more massive than $M_{\th} = 10^{14}\,M_\odot$. We quantify the differences between covariance matrices by performing a likelihood analysis with different covariance configurations, and evaluating their effect on the cosmological posteriors. To correct for the halo bias inaccuracy in the likelihood analysis, we rescale the predicted 2PCF to the mean measured one, plus the cosmology dependence from theory (see Eq.\,\ref{eq:2pcf_correction}). We constrain the parameters $\Omega_{\rm m}$ and $\sigma_8$, on which the cluster clustering is more strongly sensitive. 

The main results of our analysis can be summarized as follows.
\begin{itemize}
    \item In Sect.\,\ref{sec:binning}, we test different binning schemes, to properly extract the cosmological information. We find negligible differences when varying the width of redshift bins. We also observe a slight increase of the extracted information when increasing the number of radial bins, up to $N_{\rm r} \simeq 30$. We select the redshift bin width $\Delta z = 0.4$ and a number of radial log-spaced bins $N_{\rm r} = 30$, corresponding to $\Delta \logten (r/h^{-1}\,{\rm Mpc}) = 0.028$;
    \item In Sect.\,\ref{sec:model}, when comparing the analytical model of Eq.\,\eqref{eq:covariance} with the numerical matrix, we find that the former underestimates the covariance at intermediate and high redshift by $\sim 30\,\%$ on the diagonal and $\gtrsim 50\,\%$ on the off-diagonal terms. We ascribe this difference to the absence of high-order non-Gaussian terms and to the inaccuracy of the Poissonian shot-noise assumption, as well as the residual inaccuracy of the assumed model for the halo bias;
    \item We improve the model by adding three parameters $\{\alpha, \beta, \gamma\}$, to correct for non-Poissonian shot-noise and halo bias prediction, as well as to absorb the effect of the missing high-order terms. The parameters are fitted from simulations. We obtain an agreement within 10\,\% with the numerical matrix at all the redshifts. It should be noted that, even if the missing terms are added analytically and a perfect description of the halo bias is provided, the exact value of shot-noise cannot be predicted. Correcting the model with such fitted parameters is therefore a well-motivated procedure;
    \item From the likelihood analysis we find a difference of $\sim 40\,\%$ between the the model and numerical FoMs. Such difference drops to $\sim 5\,\%$ when adding the fitted parameters to the model. This difference is considered to be negligible in more complete analyses (e.g. richness-selected catalogs), most likely to be absorbed by the broadening of the cosmological posteriors;
    \item In Sect.\,\ref{sec:gauss_model}, we assess the relevance of the low-order non-Gaussian term, which turns out to be non-negligible at small scales, especially at high redshift;
    \item In Sect.\,\ref{sec:cosmo_dep}, we find that, in this analysis, the likelihood with cosmology-dependent covariance is statistically preferred over the iterative method. Also, we find that evaluating the covariance at a fixed wrong cosmology can lead to an under/overestimated posterior's amplitude. Moreover, neglecting the cosmology-dependence of the covariance means losing the information contained in the shot-noise term. Such information is not contained directly in the 2PCF, but is nevertheless information that characterises the clustering of clusters;
    \item  In Sect.\,\ref{sec:mass_bins}, we assess the cosmological information encoded in the shape of the halo bias by splitting our sample in mass bins, finding a significative improvement in the FoM compared to mass-threshold case. Such improvement is expected to be even more important for richness-selected halos, where this dependence can help to constrain the mass-observable relation parameters in addition to the cosmological ones.
\end{itemize}

Two main results emerge from this analysis. First, for cluster clustering a pure Gaussian model is not sufficient to correctly describe the covariance. This is due to the low number densities that characterise the spatial distribution of these objects, making non-Gaussian terms more important as the redshift and the mass threshold increase. Despite this, a simple semi-analytical model with parameters fitted from simulations permits to correct the inaccuracy of the model and gives an accurate estimate of the errors associated with the 2PCF. Although this model still requires the use of simulations to fit the covariance parameters, the number of simulations is considerably lower than the number required to compute a good numerical matrix (approximately $O(10^2)$ instead of $O(10^3)$). Furthermore, the resulting matrix is completely noise-free and accounts for the dependence on cosmological parameters.

Second, the covariance of the 2PCF contains cosmological information that is not present in the mean value. Therefore, both quantities should be taken into account in constraining cosmological parameters, to correctly extract the information enclosed in the cluster clustering, especially when the mass binning is included. Note that this may require some care when performing a combined analysis of cluster number counts and cluster clustering, as the cosmological information contained in the 2PCF covariance is also contained in the number counts. We reserve to examine this issue in detail in a future dedicated work.

In this work we show how a simple semi-analytical model can be used to accurately describe the cluster clustering covariance matrix. However, the calibration of such model is not universal, but depends on the specific properties of the survey, such as geometry or mass and redshift range. The fit of the covariance parameters must then be performed for each survey, on appropriate simulations. Moreover, such parameters may contain a non-negligible dependence on cosmology whose impact is still to be quantified. 

The natural evolution of this work involves the validation of the model for clusters selected by richness through a mass-observable relation, and with selection functions to account for the richness and redshift measurements uncertainties. Moreover, the calibration described in this work only holds for a $\Lambda$CDM Universe; further analysis is thus required to validate the model in non-standard cosmological models. Then, once a complete description of the 2PCF and its covariance will be obtained, it will be possible to exploit clustering of galaxy clusters to obtain cosmological constraints and to assess its impact in the combined analysis with other cosmological observables, at the level of accuracy that will be achieved by \Euclid. 

\begin{acknowledgements}
AF, AS, TC and SB are supported by the INFN INDARK PD51 grant. AF and AS are also supported by the ERC `ClustersXCosmo' grant agreement 716762. TC and AS are also supported by the FARE MIUR grant `ClustersXEuclid' R165SBKTMA. 
 \AckEC
\end{acknowledgements}

\bibliographystyle{aa} 
\bibliography{biblio}

\begin{thebibliography}{86}
\expandafter\ifx\csname natexlab\endcsname\relax\def\natexlab#1{#1}\fi

\bibitem[{Albrecht {et~al.}(2006)Albrecht, Bernstein, Cahn,
  {et~al.}}]{Albrecht:2006um}
Albrecht, A., Bernstein, G., Cahn, R., {et~al.} 2006, arXiv:0609591

\bibitem[{Anderson(2003)}]{Anderson:2003}
Anderson, T. 2003, An Introduction to Multivariate Statistical Analysis, Wiley
  Series in Probability and Statistics (Wiley)

\bibitem[{Angulo {et~al.}(2005)Angulo, Baugh, Frenk, Bower, Jenkins, \&
  Morris}]{Angulo:2005pt}
Angulo, R., Baugh, C.~M., Frenk, C.~S., {et~al.} 2005, \mnras, 362, L25

\bibitem[{Baldauf {et~al.}(2015)Baldauf, Mirbabayi, Simonovi\'c, \&
  Zaldarriaga}]{Baldauf:2015xfa}
Baldauf, T., Mirbabayi, M., Simonovi\'c, M., \& Zaldarriaga, M. 2015, PRD, 92,
  043514

\bibitem[{Bernstein(1994)}]{Bernstein:1993nb}
Bernstein, G.~M. 1994, \apj, 424, 569

\bibitem[{Blot {et~al.}(2020)Blot, Corasaniti, Rasera, \&
  Agarwal}]{Blot:2020cbi}
Blot, L., Corasaniti, P.-S., Rasera, Y., \& Agarwal, S. 2020, \mnras, 500, 2532

\bibitem[{Blot {et~al.}(2019)}]{Blot:2018oxk}
Blot, L. {et~al.} 2019, \mnras, 485, 2806

\bibitem[{Bond \& Myers(1996)}]{Bond:1993we}
Bond, J.~R. \& Myers, S.~T. 1996, \apj \ Suppl., 103, 1

\bibitem[{Borgani {et~al.}(1999)Borgani, Plionis, \&
  Kolokotronis}]{Borgani:1998sfa}
Borgani, S., Plionis, M., \& Kolokotronis, E. 1999, \mnras, 305, 866

\bibitem[{Bouchet {et~al.}(1995)Bouchet, Colombi, Hivon, \&
  Juszkiewicz}]{Bouchet:1994xp}
Bouchet, F.~R., Colombi, S., Hivon, E., \& Juszkiewicz, R. 1995, \aap, 296, 575

\bibitem[{Buchert(1992)}]{Buchert:1992ya}
Buchert, T. 1992, \mnras, 254, 729

\bibitem[{Buchner {et~al.}(2014)Buchner, Georgakakis, Nandra, Hsu, Rangel,
  Brightman, Merloni, Salvato, Donley, \& Kocevski}]{Buchner:2014nha}
Buchner, J., Georgakakis, A., Nandra, K., {et~al.} 2014, \aap, 564, A125

\bibitem[{Carron(2013)}]{Carron:2012pw}
Carron, J. 2013, Astron. Astrophys., 551, A88

\bibitem[{Castro {et~al.}(2020)Castro, Borgani, Dolag, Marra, Quartin, Saro, \&
  Sefusatti}]{Castro:2020yes}
Castro, T., Borgani, S., Dolag, K., {et~al.} 2020, \mnras, 500, 2316

\bibitem[{Cohn(2006)}]{Cohn:2005ex}
Cohn, J.~D. 2006, New Astron., 11, 226

\bibitem[{Colavincenzo {et~al.}(2019)}]{Colavincenzo:2018cgf}
Colavincenzo, M. {et~al.} 2019, \mnras, 482, 4883

\bibitem[{Cole {et~al.}(2005)Cole, Percival, Peacock,
  {et~al.}}]{2dFGRS:2005yhx}
Cole, S., Percival, W.~J., Peacock, J.~A., {et~al.} 2005, \mnras, 362, 505

\bibitem[{Despali {et~al.}(2016)Despali, Giocoli, Angulo, Tormen, Sheth, Baso,
  \& Moscardini}]{Despali:2015yla}
Despali, G., Giocoli, C., Angulo, R.~E., {et~al.} 2016, \mnras, 456, 2486

\bibitem[{Dodelson \& Schneider(2013)}]{Dodelson:2013uaa}
Dodelson, S. \& Schneider, M.~D. 2013, PRD, 88, 063537

\bibitem[{Eifler {et~al.}(2009)Eifler, Schneider, \& Hartlap}]{Eifler:2008gx}
Eifler, T., Schneider, P., \& Hartlap, J. 2009, \aap, 502, 721

\bibitem[{Eisenstein \& Loeb(1995)}]{Eisenstein:1994ni}
Eisenstein, D.~J. \& Loeb, A. 1995, \apj, 439, 520

\bibitem[{Eisenstein {et~al.}(2007)Eisenstein, Seo, Sirko, \&
  Spergel}]{Eisenstein:2006nk}
Eisenstein, D.~J., Seo, H.-j., Sirko, E., \& Spergel, D. 2007, \apj, 664, 675

\bibitem[{Eisenstein {et~al.}(2005)Eisenstein, Zehavi, Hoggand,
  {et~al.}}]{SDSS:2005xqv}
Eisenstein, D.~J., Zehavi, I., Hoggand, D.~W., {et~al.} 2005, \apj, 633, 560

\bibitem[{Estrada {et~al.}(2009)Estrada, Sefusatti, \&
  Frieman}]{Estrada:2008em}
Estrada, J., Sefusatti, E., \& Frieman, J.~A. 2009, \apj, 692, 265

\bibitem[{Euclid Collaboration:~Fumagalli {et~al.}(2021)Euclid
  Collaboration:~Fumagalli, Saro, Borgani, {et~al.}}]{Euclid:2021api}
Euclid Collaboration:~Fumagalli, A., Saro, A., Borgani, S., {et~al.} 2021,
  \aap, 652, A21

\bibitem[{Euclid Collaboration:~Scaramella {et~al.}(2022)Euclid
  Collaboration:~Scaramella, Amiaux, Mellier, {et~al.}}]{Euclid:2021icp}
Euclid Collaboration:~Scaramella, R., Amiaux, J., Mellier, Y., {et~al.} 2022,
  \aap, 662, A112

\bibitem[{Feldman {et~al.}(1994)Feldman, Kaiser, \& Peacock}]{Feldman:1993ky}
Feldman, H.~A., Kaiser, N., \& Peacock, J.~A. 1994, \apj, 426, 23

\bibitem[{Friedrich {et~al.}(2016)Friedrich, Seitz, Eifler, \&
  Gruen}]{Friedrich:2015nga}
Friedrich, O., Seitz, S., Eifler, T.~F., \& Gruen, D. 2016, \mnras, 456, 2662

\bibitem[{Fumagalli {et~al.}(2022)Fumagalli, Biagetti, Saro, Sefusatti, Slosar,
  Monaco, \& Veropalumbo}]{Fumagalli:2022plg}
Fumagalli, A., Biagetti, M., Saro, A., {et~al.} 2022, arXiv:2206.05191

\bibitem[{Grandis {et~al.}(2016)Grandis, Rapetti, Saro, Mohr, \&
  Dietrich}]{Grandis:2016fwl}
Grandis, S., Rapetti, D., Saro, A., Mohr, J.~J., \& Dietrich, J.~P. 2016,
  \mnras, 463, 1416

\bibitem[{Hartlap {et~al.}(2007)Hartlap, Simon, \& Schneider}]{Hartlap:2006kj}
Hartlap, J., Simon, P., \& Schneider, P. 2007, \aap, 464, 399

\bibitem[{Hu \& Kravtsov(2003)}]{Hu:2002we}
Hu, W. \& Kravtsov, A.~V. 2003, \apj, 584, 702

\bibitem[{Huetsi(2010)}]{Huetsi:2009zq}
Huetsi, G. 2010, \mnras, 401, 2477

\bibitem[{Kerscher {et~al.}(2000)Kerscher, Szapudi, \&
  Szalay}]{Kerscher:1999hc}
Kerscher, M., Szapudi, I., \& Szalay, A.~S. 2000, \apj \ Lett., 535, L13

\bibitem[{Krause \& Eifler(2017)}]{Krause:2016jvl}
Krause, E. \& Eifler, T. 2017, \mnras, 470, 2100

\bibitem[{Kravtsov \& Borgani(2012)}]{Kravtsov:2012zs}
Kravtsov, A. \& Borgani, S. 2012, AR\aap, 50, 353

\bibitem[{Lacasa \& Kunz(2017)}]{Lacasa:2017xbi}
Lacasa, F. \& Kunz, M. 2017, \aap, 604, A104

\bibitem[{Landy \& Szalay(1993)}]{Landy:1993yu}
Landy, S.~D. \& Szalay, A.~S. 1993, \apj, 412, 64

\bibitem[{Laureijs {et~al.}(2011)Laureijs, Amiaux, Arduini,
  {et~al.}}]{EUCLID:2011zbd}
Laureijs, R., Amiaux, J., Arduini, S., {et~al.} 2011, arXiv:1110.3193

\bibitem[{{Lesci} {et~al.}(2022){Lesci}, {Nanni}, {Marulli}, {Moscardini},
  {Veropalumbo}, {Maturi}, {Sereno}, {Radovich}, {Bellagamba}, {Roncarelli},
  {Bardelli}, {Castignani}, {Covone}, {Giocoli}, {Ingoglia}, \&
  {Puddu}}]{2022arXiv220307398L}
{Lesci}, G.~F., {Nanni}, L., {Marulli}, F., {et~al.} 2022, arXiv:2203.07398

\bibitem[{Lewis {et~al.}(2000)Lewis, Challinor, \& Lasenby}]{Lewis:1999bs}
Lewis, A., Challinor, A., \& Lasenby, A. 2000, \apj, 538, 473

\bibitem[{Li {et~al.}(2019)Li, Singh, Yu, Feng, \& Seljak}]{Li:2018scc}
Li, Y., Singh, S., Yu, B., Feng, Y., \& Seljak, U. 2019, JCAP, 01, 016

\bibitem[{Lippich {et~al.}(2019)}]{Lippich:2018wrx}
Lippich, M. {et~al.} 2019, \mnras, 482, 1786

\bibitem[{Majumdar \& Mohr(2004)}]{Majumdar:2003mw}
Majumdar, S. \& Mohr, J.~J. 2004, \apj, 613, 41

\bibitem[{Mana {et~al.}(2013)Mana, Giannantonio, Weller, Hoyle, Huetsi, \&
  Sartoris}]{Mana:2013qba}
Mana, A., Giannantonio, T., Weller, J., {et~al.} 2013, \mnras, 434, 684

\bibitem[{Manera {et~al.}(2010)Manera, Sheth, \& Scoccimarro}]{Manera:2009ak}
Manera, M., Sheth, R.~K., \& Scoccimarro, R. 2010, \mnras, 402, 589

\bibitem[{Marulli {et~al.}(2021)Marulli, Veropalumbo, Garc\'\i{}a-Farieta,
  Moresco, Moscardini, \& Cimatti}]{Marulli:2020uyy}
Marulli, F., Veropalumbo, A., Garc\'\i{}a-Farieta, J.~E., {et~al.} 2021, \apj,
  920, 13

\bibitem[{Marulli {et~al.}(2016)Marulli, Veropalumbo, \&
  Moresco}]{Marulli:2015jil}
Marulli, F., Veropalumbo, A., \& Moresco, M. 2016, Astron. Comput., 14, 35

\bibitem[{Marulli {et~al.}(2018)Marulli, Veropalumbo, Sereno,
  {et~al.}}]{Marulli:2018owk}
Marulli, F., Veropalumbo, A., Sereno, M., {et~al.} 2018, \aap, 620, A1

\bibitem[{Meiksin \& White(1999)}]{Meiksin:1998mu}
Meiksin, A. \& White, M.~J. 1999, \mnras, 308, 1179

\bibitem[{Miller {et~al.}(2001)Miller, Nichol, \& Batuski}]{Miller:2001cf}
Miller, C.~J., Nichol, R.~C., \& Batuski, D.~J. 2001, \apj, 555, 68

\bibitem[{Mo \& White(1996)}]{Mo:1995cs}
Mo, H.~J. \& White, S. D.~M. 1996, \mnras, 282, 347

\bibitem[{Mohammad \& Percival(2021)}]{Mohammad:2021aqc}
Mohammad, F.~G. \& Percival, W.~J. 2021, arXiv:2109.07071

\bibitem[{Monaco(2016)}]{Monaco:2016pys}
Monaco, P. 2016, Galaxies, 4, 53

\bibitem[{Monaco {et~al.}(2002)Monaco, Theuns, \& Taffoni}]{Monaco:2001jg}
Monaco, P., Theuns, T., \& Taffoni, G. 2002, \mnras, 331, 587

\bibitem[{Moresco {et~al.}(2021)Moresco, Veropalumbo, Marulli, Moscardini, \&
  Cimatti}]{Moresco:2020quj}
Moresco, M., Veropalumbo, A., Marulli, F., Moscardini, L., \& Cimatti, A. 2021,
  \apj, 919, 144

\bibitem[{Morrison \& Schneider(2013)}]{Morrison:2013tqa}
Morrison, C.~B. \& Schneider, M.~D. 2013, JCAP, 11, 009

\bibitem[{Moscardini {et~al.}(2000)Moscardini, Matarrese, Lucchin, \&
  Rosati}]{Moscardini:1999ba}
Moscardini, L., Matarrese, S., Lucchin, F., \& Rosati, P. 2000, \mnras, 316,
  283

\bibitem[{Moutarde {et~al.}(1991)Moutarde, Alimi, Bouchet, Pellat, \&
  Ramani}]{Moutarde:1991evx}
Moutarde, F., Alimi, J.~M., Bouchet, F.~R., Pellat, R., \& Ramani, A. 1991,
  \apj, 382

\bibitem[{Munari {et~al.}(2017)Munari, Monaco, Sefusatti, Castorina, Mohammad,
  Anselmi, \& Borgani}]{Munari:2016aut}
Munari, E., Monaco, P., Sefusatti, E., {et~al.} 2017, \mnras, 465, 4658

\bibitem[{Norberg {et~al.}(2009)Norberg, Baugh, Gaztanaga, \&
  Croton}]{Norberg:2008tg}
Norberg, P., Baugh, C.~M., Gaztanaga, E., \& Croton, D.~J. 2009, \mnras, 396,
  19

\bibitem[{O'Connell {et~al.}(2016)O'Connell, Eisenstein, Vargas, Ho, \&
  Padmanabhan}]{OConnell:2015src}
O'Connell, R., Eisenstein, D., Vargas, M., Ho, S., \& Padmanabhan, N. 2016,
  \mnras, 462, 2681

\bibitem[{Paech {et~al.}(2017)Paech, Hamaus, Hoyle, Costanzi, Giannantonio,
  Hagstotz, Sauerwein, \& Weller}]{Paech:2016hod}
Paech, K., Hamaus, N., Hoyle, B., {et~al.} 2017, \mnras, 470, 2566

\bibitem[{Percival {et~al.}(2022)Percival, Friedrich, Sellentin, \&
  Heavens}]{Percival:2021cuq}
Percival, W.~J., Friedrich, O., Sellentin, E., \& Heavens, A. 2022, \mnras,
  510, 3207

\bibitem[{Philcox \& Eisenstein(2019)}]{Philcox:2019xzt}
Philcox, O. H.~E. \& Eisenstein, D.~J. 2019, \mnras, 490, 5931

\bibitem[{{Planck Collaboration VI.}(2020)}]{Planck:2018vyg}
{Planck Collaboration VI.} 2020, A\&A, 641, A6, [Erratum: A\&A 652, C4 (2021)]

\bibitem[{{Planck Collaboration XVI.}(2014)}]{Planck:2013pxb}
{Planck Collaboration XVI.} 2014, A\&A, 571, A16

\bibitem[{Pope \& Szapudi(2008)}]{Pope:2007vz}
Pope, A.~C. \& Szapudi, I. 2008, \mnras, 389, 766

\bibitem[{Pratt {et~al.}(2019)Pratt, Arnaud, Biviano, Eckert, Ettori, Nagai,
  Okabe, \& Reiprich}]{Pratt:2019cnf}
Pratt, G.~W., Arnaud, M., Biviano, A., {et~al.} 2019, Space Sci. Rev., 215, 25

\bibitem[{Sahni \& Coles(1995)}]{Sahni:1995rm}
Sahni, V. \& Coles, P. 1995, Phys. Rept., 262, 1

\bibitem[{Sartoris {et~al.}(2016)Sartoris, Biviano, Fedeli,
  {et~al.}}]{Sartoris:2015aga}
Sartoris, B., Biviano, A., Fedeli, C., {et~al.} 2016, \mnras, 459, 1764

\bibitem[{Schuecker {et~al.}(2003)Schuecker, Bohringer, Collins, \&
  Guzzo}]{Schuecker:2002ti}
Schuecker, P., Bohringer, H., Collins, C.~A., \& Guzzo, L. 2003, \aap, 398, 867

\bibitem[{Scoccimarro {et~al.}(1999)Scoccimarro, Zaldarriaga, \&
  Hui}]{Scoccimarro:1999kp}
Scoccimarro, R., Zaldarriaga, M., \& Hui, L. 1999, \apj, 527, 1

\bibitem[{Sellentin \& Heavens(2016)}]{Sellentin:2015waz}
Sellentin, E. \& Heavens, A.~F. 2016, \mnras, 456, L132

\bibitem[{Senatore \& Zaldarriaga(2015)}]{Senatore:2014via}
Senatore, L. \& Zaldarriaga, M. 2015, JCAP, 02, 013

\bibitem[{Sereno {et~al.}(2015)Sereno, Veropalumbo, Marulli, Covone,
  Moscardini, \& Cimatti}]{Sereno:2014eea}
Sereno, M., Veropalumbo, A., Marulli, F., {et~al.} 2015, \mnras, 449, 4147

\bibitem[{Spiegelhalter {et~al.}(2002)Spiegelhalter, Best, Carlin, \& van~der
  Linde}]{Spiegelhalter:2002yvw}
Spiegelhalter, D.~J., Best, N.~G., Carlin, B.~P., \& van~der Linde, A. 2002, J.
  Roy. Statist. Soc. B, 64, 583

\bibitem[{Takada \& Hu(2013)}]{Takada:2013wfa}
Takada, M. \& Hu, W. 2013, PRD, 87, 123504

\bibitem[{Taylor {et~al.}(2013)Taylor, Joachimi, \& Kitching}]{Taylor:2012kz}
Taylor, A., Joachimi, B., \& Kitching, T. 2013, \mnras, 432, 1928

\bibitem[{Tinker {et~al.}(2010)Tinker, Robertson, Kravtsov, Klypin, Warren,
  Yepes, \& Gottlober}]{Tinker:2010my}
Tinker, J.~L., Robertson, B.~E., Kravtsov, A.~V., {et~al.} 2010, \apj, 724, 878

\bibitem[{To {et~al.}(2021{\natexlab{a}})To, Krause, Rozo,
  {et~al.}}]{DES:2020mlx}
To, C., Krause, E., Rozo, E., {et~al.} 2021{\natexlab{a}}, PRL, 126, 141301

\bibitem[{To {et~al.}(2021{\natexlab{b}})}]{DES:2020uce}
To, C.-H. {et~al.} 2021{\natexlab{b}}, \mnras, 502, 4093

\bibitem[{Valageas {et~al.}(2011)Valageas, Clerc, Pacaud, \&
  Pierre}]{Valageas:2011mz}
Valageas, P., Clerc, N., Pacaud, F., \& Pierre, M. 2011, \aap, 536, A95

\bibitem[{Veropalumbo {et~al.}(2014)Veropalumbo, Marulli, Moscardini, Moresco,
  \& Cimatti}]{Veropalumbo:2013cua}
Veropalumbo, A., Marulli, F., Moscardini, L., Moresco, M., \& Cimatti, A. 2014,
  \mnras, 442, 3275

\bibitem[{Wadekar \& Scoccimarro(2020)}]{Wadekar:2019rdu}
Wadekar, D. \& Scoccimarro, R. 2020, PRD, 102, 123517

\bibitem[{Xu {et~al.}(2012)Xu, Padmanabhan, Eisenstein, Mehta, \&
  Cuesta}]{Xu:2012hg}
Xu, X., Padmanabhan, N., Eisenstein, D.~J., Mehta, K.~T., \& Cuesta, A.~J.
  2012, \mnras, 427, 2146

\end{thebibliography}

\begin{appendix}
\section{Non-Poissonian shot-noise}\label{app:shot_noise}

To better assess the impact of shot-noise and its deviation from the Poissonian prediction, we study the power spectrum, i.e. the quantity that is directly affected by this correction. To avoid complications due to the redshift integration and the geometry of the survey, we measure the power spectrum from 1000 cubical boxes of size $L = 3870\,h^{-1}\,$Mpc, with the same properties and cosmology of the light cones described in Sect.\,\ref{sec:simulations}. We consider three redshifts $z = 0.5, 1.0, 1.5$. 

We compute the analytical total halo power spectrum as
\begin{equation}
    P_{\rm h,tot}(k) = \overline{b}^2\,P_{\rm m}(k) + \frac{1}{\overline{n}}\,,
\end{equation}
where the matter power spectrum  is calculated by means of the  \texttt{CAMB} code \citep{Lewis:1999bs}. We compare such quantity with the measured total power spectrum averaged over the 1000 boxes, for the three redshift values. 

\begin{figure}[h]
    \centering
    \includegraphics[width=0.49\textwidth]{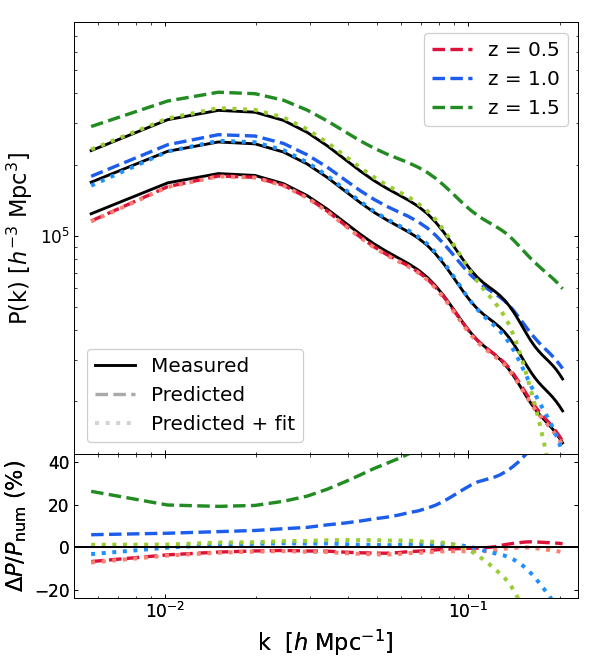}
    \caption{Halo power spectrum. \textit{Top panel:} measured (black solid lines), predicted (darker dashed lines) and fitted (lighter dotted lines) halo power spectrum for boxes at three different redshifts. \textit{Bottom panel:} percent residuals of the predictions with respect to the measured one.}
    \label{fig:pk}
\end{figure}

In Fig.\,\ref{fig:pk} we show the comparison of the measured and predicted power spectra (for a better comparison, we show the halo power spectrum, i.e. the total one minus the shot-noise): while at low redshift the two quantities are in agreement on almost all the scales, we can observe a clear deviation between the observed and predicted spectra that increases with redshift, up to more than 20\,\%. We try to correct these discrepancies by fitting the two parameters $\{\alpha,\beta\}$, which account respectively for the correction to shot-noise and halo bias, directly from the power spectrum; the best-fit are shown in Table\,\ref{tab:fit_pk} and the resulting power spectra are shown in Fig\,\ref{fig:pk} (dotted lines). We obtain an agreement of the fitted power spectra within 5 per cent level at all the linear scales, at all redshifts. Nevertheless, the values deviate by several $\sigma$ from the best-fit parameters found from the covariance fit in the corresponding redshift intervals (see Table\,\ref{table:params_fit}). This confirms that the parameters in the covariance, in addition to correct for the wrong prediction of bias and shot-noise, also absorb the effect of the missing higher-order terms in the model.

\begin{table}[h]
\centering           
\caption{Best-fit parameters for the power spectrum parameters.}
\begin{tabular}{l c c}       
\hline  
Redshift & $\alpha$ & $\beta$ \\
\hline
0.5  & 0.012 $\pm$ 0.010 & 0.987 $\pm$ 0.002 \\
1.0  & 0.114 $\pm$ 0.004 & 1.006 $\pm$ 0.002 \\
1.5  & 0.104 $\pm$ 0.002 & 1.013 $\pm$ 0.002 \\
\hline
Reference &  0 & 1 \\
\hline
\label{tab:fit_pk} 
\end{tabular}
\end{table}

\section{Covariance parameters fit} \label{app:cov_fit}

\begin{figure}
    \centering
    \includegraphics[width=0.49\textwidth]{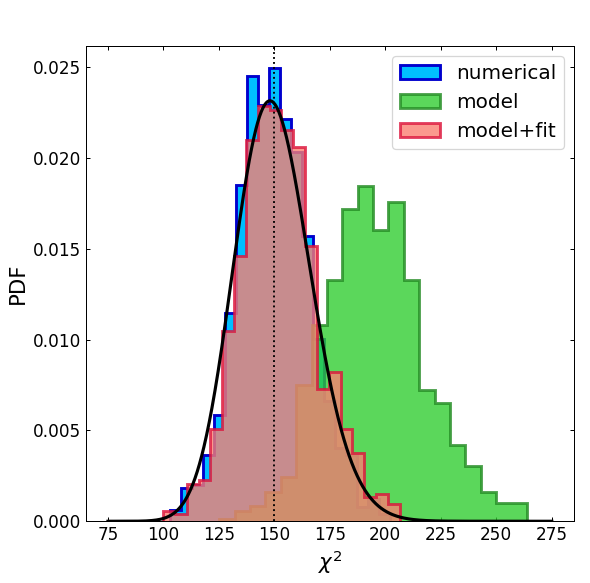}
    \caption{$\chi^2$ distribution for the numerical, analytical and analytical with fitted parameters covariance matrices. Reference distribution in black.}
    \label{fig:chi2}
\end{figure}

We show here the results of the covariance parameters fit, following the method proposed in \citet{Fumagalli:2022plg}.
In Fig.\,\ref{fig:chi2} we show the $\chi^2$ distributions computed with respect to the measurements from the 1000 lightcones, for the three covariance matrices: numerical covariance in blue, analytical covariance in green and analytical covariance with fitted parameters in red. In Table\,\ref{tab:chi2} we report the mean and standard deviation for each distribution, with the corresponding $1\sigma$ uncertainties computed with the bootstrap technique. By construction, we expect the numerical matrix to perfectly follow the reference distribution. While this is true for the mean value, the variance differs by $\sim 4\sigma$ from the expected value; such discrepancy is ascribed to the noise in the numerical matrix that tightens the distribution. Anyway, since the errors are quite small, this distribution can be considered in good agreement with the expected one. Instead, the model of Eq.\,\eqref{eq:covariance} produces a distribution that is several $\sigma$ off the expectation, confirming that this model is not suitable for describing the covariance of data. Finally, the fitted matrix turns out to be in good agreement with the reference distribution, both for the mean value and the variance. This proves the goodness of our fit and ensures that the resulting model is able to correctly describe the covariance as well as, if not better than, the numerical matrix. The fitting process provides consistent results when the fit is performed with only 100 simulations.

\begin{table}
\centering           
\caption{$\chi^2$distribution test values.}
\begin{tabular}{l c c}       
\hline  
 & mean & variance \\
\hline
Numerical   & 149.0 $\pm$ 0.5 & 255.0 $\pm$ 11.7 \\
Model       & 195.0 $\pm$ 0.7 & 494.0 $\pm$ 22.8 \\
Model + fit & 152.0 $\pm$ 0.6 & 302.0 $\pm$ 13.9 \\
\hline
Reference &  150 & 300 \\
\hline
\label{tab:chi2} 
\end{tabular}
\end{table}

\section{Cosmology-dependent matrix} \label{app:cosmo_mocks}
To further explore the comparison between the fully cosmology-dependent likelihood analysis and the iterative method, we generate 100 synthetic light cones starting from a Gaussian distribution, with amplitude given by the covariance model at the input cosmology. In this way, we ensure that the Gaussian distribution is the true likelihood describing the data, and not just an approximation. We repeat the analysis of the 100 light cones described in Sect.\,\ref{sec:cosmo_dep}, finding a mean value $\langle \Delta {\rm DIC} \rangle_{\rm synth} = -13.7 \pm 2.1$, to be compared with the value from the analysis of the 100 PINOCCHIO mocks, i.e., $\langle \Delta {\rm DIC} \rangle_{\rm sims}= -11.5 \pm 1.6$. Moreover, by comparing the FoM of the cosmology dependent covariance and the iterative method, we obtain a mean variation of $\langle \Delta {\rm FoM} \rangle_{\rm synth} = 176 \pm 38 \,\%$ for the synthetic catalogs, and $\langle \Delta {\rm FoM} \rangle_{\rm sims} = 142 \pm 33 \,\%$ for the PINOCCHIO mocks. Both the DIC and the FoM analyses indicate that the two analysis are fully consistent. Although this result still does not define which posteriors are correct in case the true likelihood is unknown, it shows that at least for this particular analysis the narrowing of the posteriors does not primarily depends on some wrong approximation of the likelihood function. In other words, when a Gaussian likelihood is assumed, it is possible to actually extract information from the cosmology dependence of the covariance. 

\section{Cosmology dependence of number counts covariance}\label{app:numbercounts}
\begin{figure}[h]
    \centering
    \includegraphics[width=0.49\textwidth]{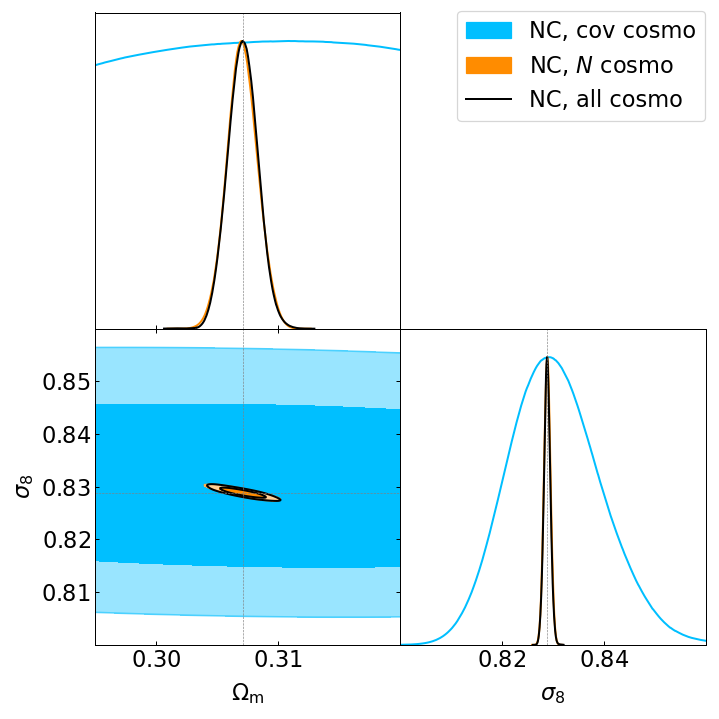}
    \caption{Same of Fig.\,\ref{fig:post_cosmo} for cluster number counts.}
    \label{fig:post_cosmo_nc}
\end{figure}

We discuss here some additional considerations about the cosmology-dependence of the covariance, using as an example the cluster number counts. It has been shown in \citet{Euclid:2021api} that the use of a fixed covariance in the likelihood analysis can bring an under/overestimation of the FoM by more than 40\,\%, if the cosmology at which the covariance is evaluated deviates from the fiducial values of an amount of 2$\sigma$ from \citet{Planck:2018vyg}. Instead, there are not significant differences between the input and varying covariance cases.  We perform the same test on the cosmology-dependence of the covariance described in Sect.\,\ref{sec:cosmo_dep}, obtaining the posteriors shown in Fig.\,\ref{fig:post_cosmo_nc}. Unlike for the clustering case, for number counts the mean value is much more constraining than its covariance, making the degeneracy direction of the latter totally irrelevant. In this case, thus, the covariance only contributes as an estimation of the uncertainty, without adding further independent information.

\end{appendix}
\end{document}